\documentclass[3p,onecolumn,times,review]{elsarticle}
\usepackage{dcolumn}
\usepackage[T1]{fontenc}
\usepackage{lmodern}
\usepackage[english]{babel}
\usepackage{amsmath,amsfonts,amsthm,amssymb,amsbsy,wasysym,nicefrac}
\usepackage{mathtools, cuted}
\usepackage{braket}
\usepackage{bm}
\usepackage{booktabs}
\usepackage[labelfont=bf,justification=justified ,singlelinecheck=false,font=footnotesize]{caption}
\usepackage{lineno}
\usepackage{adjustbox}
\usepackage[para,flushleft]{threeparttable}

\newcommand\nodata{ ~$\cdots$~ } 
\newcommand\etal{~\textit{et al.}~} 
\journal{Journal of Molecular Spectroscopy}

\usepackage[final, authormarkuptext=name]{changes}

\definechangesauthor[name={A}, color=red]{A}

\usepackage[colorlinks, linkcolor = black, citecolor = black, filecolor = black, urlcolor = blue,hyperfootnotes, bookmarks]{hyperref}
\begin{document}

\begin{frontmatter}

\title{Mass-independent analysis of the stable isotopologues of gas-phase titanium monoxide -- TiO}

\author[kassel]{Alexander A. Breier\corref{mycorrespondingauthor}}
\ead{a.breier@physik.uni-kassel.de}
\author[kassel]{Bj\"{o}rn Wa\ss muth}
\author[kassel]{Guido W. Fuchs}
\author[mainz]{J\"{u}rgen Gauss}
\author[kassel]{Thomas F. Giesen}

\address[kassel]{Laboratory for Astrophysics, Institute of Physics, University of Kassel, 34132 Kassel, Germany}
\address[mainz]{Institut f\"{u}r Physikalische Chemie, Universit\"{a}t Mainz, 55099 Mainz, Germany}
\cortext[mycorrespondingauthor]{Corresponding author}

\begin{abstract}
More than 130 pure rotational transitions of $^{46}$TiO, $^{47}$TiO, $^{48}$TiO, $^{49}$TiO, $^{50}$TiO, and $^{48}$Ti$^{18}$O are recorded using a high-resolution mm-wave supersonic jet spectrometer in combination with a laser ablation source. For the first time a mass-independent Dunham-like analysis is performed encompassing rare titanium monoxide isotopologues, and are compared to results from high-accuracy quantum-chemical calculations. The obtained parametrization reveals for titanium monoxide effects due to deviations from the Born-Oppenheimer approximation. Additionally, the dominant titanium properties enable an insight into the electronic structure of TiO by analyzing its hyperfine interactions. Further, based on the mass-independent analysis, the frequency positions of the pure rotational transitions of the short lived rare isotopologue $^{44}$TiO are predicted with high accuracy, i.e., on a sub-MHz uncertainty level. This allows for dedicated radio-astronomical searches of this species in core-collapse environments of supernovae.
\end{abstract}


\end{frontmatter}

%\linenumbers

\section{Introduction}
Since the beginning of the 20\,th's century TiO (titanium monoxide) belongs to one of the best studied diatomic molecules of astrophysical relevance \cite{Fowler1904,Merrill1962}.  The molecule is routinely used to characterize circumstellar environments \cite{Kaminski2013,Kervella2016,Kaminski2017} and to classify stars within the MK spectral classification scheme \cite{Morgan1973,Tsuji1986,Schiavon1999}. Recently, TiO has been also observed in hot exoplanet atmospheres \cite{Fortney2008,Evans2016,Tennyson2016,Sedaghati2017}.

The rare and non-stable titanium isotope $^{44}$Ti is of general interest for the understanding of processes that take place in core-collapse supernovae, where it is thought to be synthesized in significant quantities \cite{Timmes1996,Lee2017}. 
In principle, the $\gamma$-ray emission from the decay of $^{44}$Ti nuclei can be used as a tool for the search of remnants of recent supernovae (less than approx 1,000 years old) \cite{Iyudin1998}.  
However, an unambiguous assignment of  the X- and $\gamma-$ray emission of $^{44}$Ti to specific sources has been so far only possible in the case of the supernovae remnant (SNR) Cassiopeia A using the IBIS/ISGRI instrument on board of the INTEGRAL satellite \cite{Tsygankov2016, Wongwathanarat2016a} and for SN 1987A using the NuSTAR space observatory \cite{Boggs2015}.

An alternative approach to identify the rare isotope in these environments could be the detection of the diatomic molecule $^{44}$TiO at radio wavelengths where modern large scale facilities, like ALMA (Atacama Large Millimeter/submillimeter Array), allow for high spatial resolution and high sensitivity. The short lifetime of $^{44}$Ti of 58.9$\pm$0.3 a \cite{Ahmad2006} hinders an easy laboratory approach to  the accurate determination of the rotational transitions of this molecule. However, mass-independent studies of titanium monoxides can be used to accurately predict the line positions. 

Observations of $^{44}$TiO and other more easily to detect stable isotopologues could result in valuable information about the fractional abundance of the titanium isotopes as result of core-collapse dynamics. \\   
In this work a mass-independent study has been performed using the Dunham approach \cite{Dunham1932, VanVleck1936} which besides its astronomical interest concerning the $^{44}$TiO is also of fundamental spectroscopic interest. The use of Dunham coefficients is a powerful tool that allows to obtain a comprehensive picture of diatomic molecules independent of the individual isotope composition. It includes aspects of the potential anharmonicity, the interaction between vibrational and electronic motion and effects of the Born-Oppenheimer breakdown and thus combines many aspects of molecular motion which results in high-accuracy parameters. 
In addition, the structure of TiO can be determined purely from experimental data by this method. 
For this to work many data from previous experiments is combined starting with publications from the late 1920, e.g., the first assigned electronic TiO $\alpha$ system ($C^3\Delta - X^3\Delta)$ \cite{Christy1929}, until recent work by Lincowski\etal\cite{Lincowski2016} about rotational transitions of several titanium isotopologues. There exist five stable Ti isotopes, namely  $^{46-50}$Ti  and three oxygen isotopes $^{16-18}$O. In this work six out of the fifteen possible combinations have been investigated ($^{46-50}$Ti$^{16}$O and $^{48}$Ti$^{18}$O).

\section{Titanium monoxide}
A rich literature is available about the rovibronic energy levels of the main isotopologue $^{48}$Ti$^{16}$O which has been summarized by McKemmish\etal\cite{McKemmish2017}. In the here presented work the laser-induced fluorescence (LIF) studies of the $\gamma^\prime$ system $(B^3\Pi-X^3\Delta)$\cite{Amiot1995} and $\gamma$  $(A^3\Phi-X^3\Delta)$  \cite{Ram1999} are of particular interest. 

The $^{47}$TiO isotopologue was observed by Fletcher\etal\cite{Fletcher1993} using the high-resolution molecular beam LIF spectra from the $\gamma^\prime$(0,0) band and the  ground-state hyperfine parameters of $^{47}$Ti (I=$\nicefrac{5}{2}$) were determined by using combination difference analysis. 
In the work of Barnes\etal\cite{Barnes1996} all isotopologues were observed in the $\gamma$(0,0) system by their LIF signals from a double-resonance experiment of rotationally jet-cooled TiO. 
More recently the $\gamma^\prime$(1,0) band of $^{46}$TiO was investigated by analysis of the LIF spectra by Amiot\etal\cite{Amiot2002}. In the same year Kobayashi\etal\cite{Kobayashi2002} observed all stable isotopologues by measuring the $(E^3\Pi-X^3\Delta)$ transitions.
\\
For the main isotopologue $^{48}$TiO purely high-resolution rotational data of the ground state were investigated by Namiki\etal\cite{Namiki1998} and Kania\etal\cite{Kania2008}. 
Recently, Lincowski\etal\cite{Lincowski2016} presented rotational measurements of all stable titanium isotopologues. 

\begin{figure}[ht]
	\begin{center}
			\includegraphics[width=0.49\textwidth]{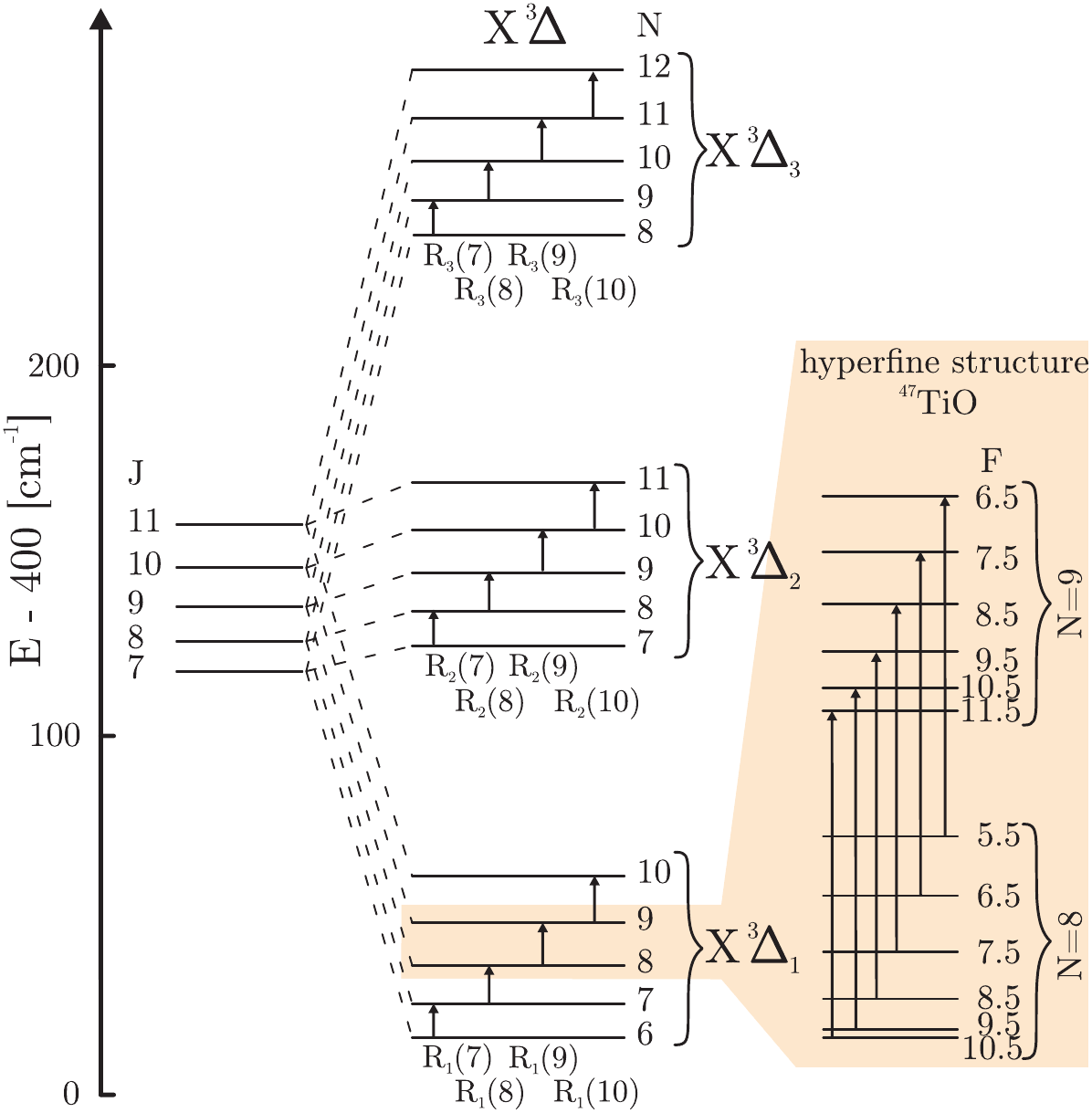}
		\caption{Fine and hyperfine structure of (X$^3\Delta$) TiO. The spin-orbit components are separated by 100\,cm$^{-1}$. For $^{47}$TiO the hyperfine splitting of the X$^3\Delta_1$, $N=9\leftarrow N=8$, and $\Delta F = + 1$ transitions are depicted.\label{fig:nrg_lvl}}
	\end{center}
\end{figure}

From previous works \cite{Amiot1995,Fletcher1993} it can be concluded that the electronic ground state  X$^3\Delta$  is well represented by a $\dots(8\sigma)^2(3\pi)^4(9\sigma)^1(1\delta)^1$ electron configuration with two unpaired electrons occupying the 9$\sigma$ and 1$\delta$ orbital. The $(9\sigma)^1$ molecular orbital (MO) originates mainly from the atomic Ti $4s$ and the MO $(1\delta)^1$ comes from a pure atomic Ti $3d$ orbital. This makes TiO the simplest molecule with a $3d$ orbital used in a molecular bonding \cite{Merer1989}. Furthermore, the two unpaired electrons are predicted to strongly polarize the molecule \cite{Fletcher1993, Bridgeman2000}  resulting in a $X^3\Delta$ state as lowest electronic configuration. 
This leads to a coupling of the electronic $(\Lambda=2)$ and the spin ($\Sigma=1$) angular momentum according to Hund's case~$(a)$ resulting in three energetically well separated sub states  $^{3}\Delta_{1}$,   $^{3}\Delta_{2}$ and  $^{3}\Delta_{3}$ (adapting the usual nomenclature $^{2\Sigma+1}\Lambda_{\Omega}$ with $\Omega=\Lambda-\Sigma,\cdots, \Lambda+\Sigma$), see Fig. \ref{fig:nrg_lvl}. 
The analysis of optical Stark spectra shows a permanent electrical dipole moment of 3.34(1)\,D for the ground state $X^3\Delta_1$\cite{Steimle2003}.

\section{Experiment}
High-resolution sub-mm--wavelength absorption spectra of 132 TiO rotational transitions in the ground state (X$^3\Delta$) have been recorded using the Supersonic Jet Spectrometer for Terahertz Applications (SuJeSTA). The experimental setup has been described in Ref.~\cite{Breier2016}.% (see, for example, Fig. 2 therein).  

In brief, a 1064\,nm intense Q-switched Nd:YAG  laser beam at 30\,Hz repetition rate is focused on a  solid titanium rod (99,6\% purity, Goodfellow).  The ablated material is seeded in either a pulsed gas flow of a 5\% N$_2$O or a 1.25\% O$_2$ mixture in helium. 
For measurements on the Ti$^{18}$O isotopologue an admixture of $^{18}$O$_2$  (Campro Scientific GmbH, 97 atom \%) is used with a mixing ratio $^{18}$O$_2$:$^{16}$O$_2$ of 1:5, \added[id=A]{ which allows to control the stability of molecular production by checking to the signal of the most abundant TiO isotopologue}.
The gas at room temperature has a 2\,bar stagnation pressure at the injecting valve but before reaching the titanium rod, it undergoes adiabatic pre-cooling to below 100~K. 

\begin{figure*}[h]
	\begin{center}
		\resizebox{.995\linewidth}{!}{
			\includegraphics{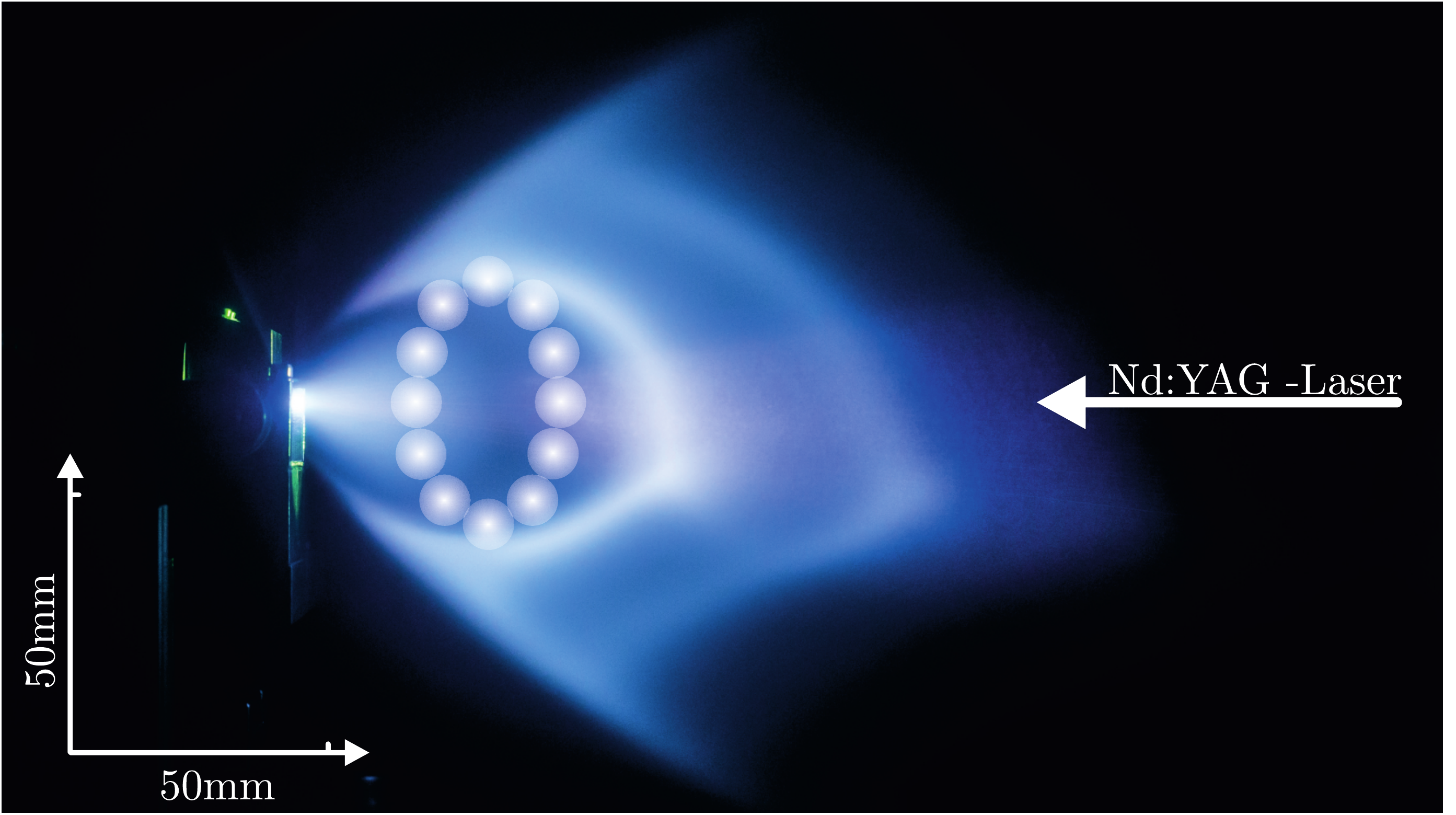}
		}
		\caption{Photograph of a supersonic jet (Exposure time: 2.9\,s, f5.6, ISO 3200) superimposed with a sketch of a multi-reflection submm-wavelength beam (300\,GHz). The jet contains titanium vapor from a laser ablation, seeded in argon buffer gas (2\,bar).\label{fig:jet}}
	\end{center}
\end{figure*}

During a few $\mu$sec the ablated titanium can chemically react with the gas mixture in a $\approx$100 mm$^3$ large reaction channel at a few hundred mbar pressure before adiabatically expanding  into a vacuum chamber where it can be observed as supersonic jet (see \cite{Neubauer2003} for more details on the ablation source). Fig.~\ref{fig:jet} shows  a supersonic jet expansion with well pronounced shock fronts. To enhance the contrast of the photograph we used argon instead of helium which emits intense blue light from electronically excited states. The TiO molecules seeded in the adiabatically cooled helium jet interact with a submm-wavelength probe beam 20\,mm down-stream from the nozzle exit, where they have a rotational temperature of a few tens Kelvin. 

A multi-pass optics (12 paths Herriott type) perpendicular to the jet propagation is used to enhance the absorption. 
Radiation between 250\,to\,385\,GHz is produced by a cw-operating cascaded multiplier chain (AMC-9 + triple stage from Virginia Diodes, VDI) where the 9\,to\,14\,GHz output signal of a tunable synthesizer (VDI) is amplified and frequency multiplied by a factor of 27. 
The signal is detected by a liquid-He cooled InSb hot-electron bolometer (QMC instruments) and data is recorded during a 100\,$\mu$s time window. A low-noise amplifier and band-pass filter (SR560, Scientific Instruments) is used to reduce the low frequency noise of the signal before storing the time-dependent signal on a computer. 
A 0.1\,MHz step-scan modus is used to scan over spectral ranges of typically 40-50 MHz, to cover individual rotational TiO lines, and their hyperfine components. To increase the signal-to-noise ratio the signal at each frequency position is averaged over eight laser shots.  
The accuracy of the measured line center positions is 10\,kHz with typical line widths (FWHM) of 0.6\,MHz.\\

\section{Measurements and Data Reduction} 
We measured 132 ground-state rotational absorption transitions of TiO in natural abundance \cite{Shima1993,IUPAC2013} and assigned 14 lines to the main isotopologue $^{48}$TiO\footnote{If not otherwise stated O refers to the $^{16}$O isotope.} (73.72\%) and 13 lines to each of the less abundant isotopologues $^{46}$TiO (8.25\%) and $^{50}$TiO  (5.18\%) (see Tab.~ \ref{tab:measurment_even}). 
Although, the rotational temperature in the jet is low, we were able to measure all spin-orbit components ($\Omega$ = 1, 2, and 3) with a signal-to-noise ratio of better than 5. 
Species of the odd numbered nuclei, $^{47}$Ti and $^{49}$Ti, exhibit prominent hyperfine spectra due to nuclear spins of I($^{47}$Ti)=$\nicefrac{5}{2}$ and I($^{49}$Ti) = $\nicefrac{7}{2}$, respectively. 
We measured 47 $^{47}$TiO (7.44\%) transitions encompassing five lines of the $^3\Delta_1$ spin-orbit component and three $^3\Delta_2$ lines each with six hyperfine components (except one  weak line). 
For $^{49}$TiO (5.41\%) 40 transitions of the $^3\Delta_1$ component were recorded encompassing five rotational transitions with eight hyperfine splittings each (see Tab.~\ref{tab:measurment_odd}). 
Furthermore, we recorded five rotational transitions of the $^3\Delta_1$ state of $^{48}$Ti$^{18}$O (see Tab.\ref{tab:measurment_18even}).
For the main isotopologue $^{48}$TiO (X$^3\Delta_2$) two rotational transitions in their first vibrationally excited state ($\nu$=1) have been observed. In addition, we observed eight not yet measured titanium dioxide (TiO$_2$) lines, listed in Tab.~\ref{tab:measurment_tio2}.\\

\begin{figure*}[h]
	\begin{center}
		\resizebox{.995\linewidth}{!}{
			\includegraphics{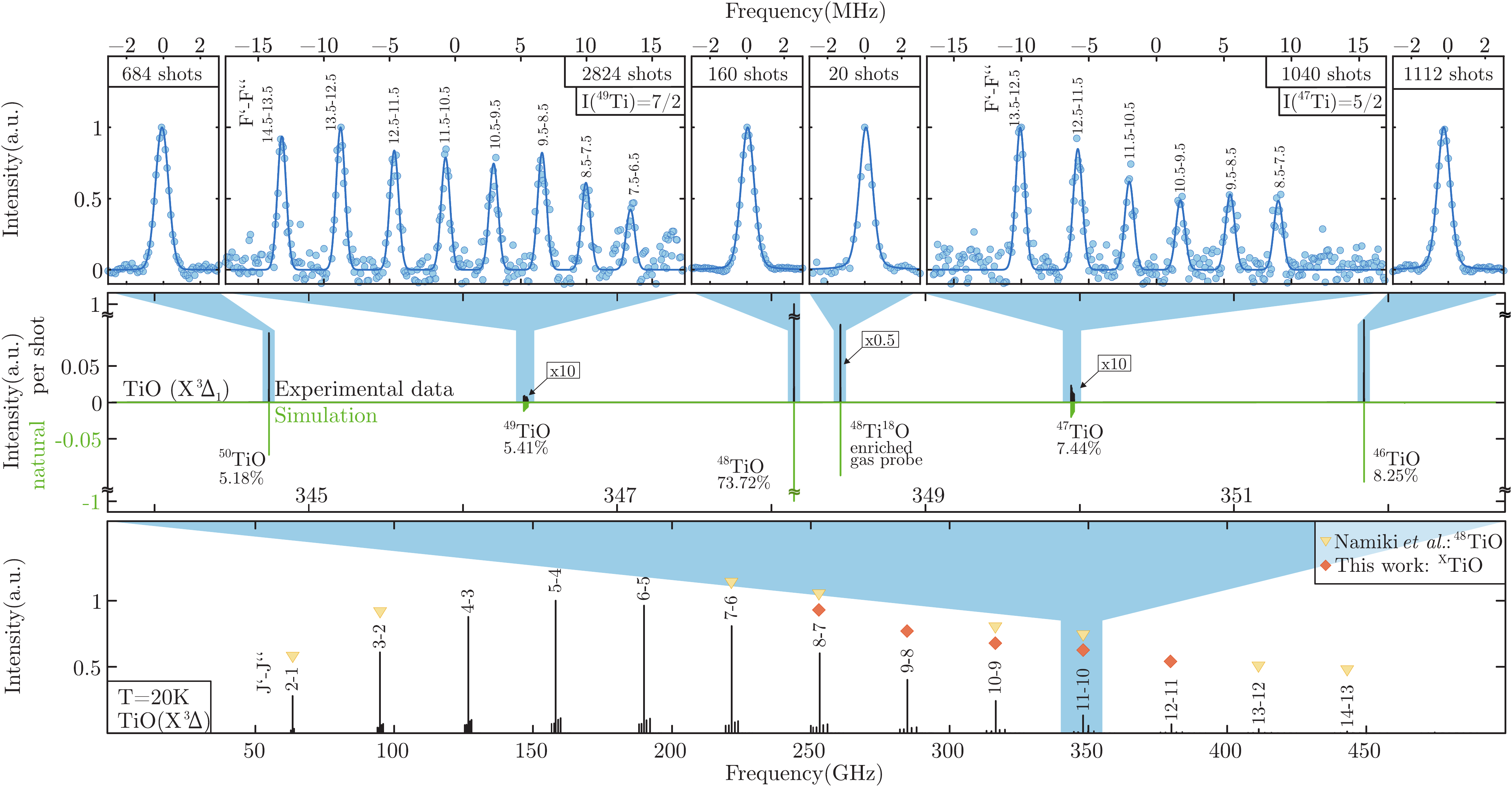}
		}
		\caption{The bottom frame shows simulated stick spectra of titanium monoxide in the rotational ground state, $X^3\Delta$, at a temperature of 20\,K and frequencies up to 500\,GHz. Triangles mark the previously measured rotational transitions of the main isotopologue $^{48}$TiO by Namiki\etal \cite{Namiki1998}. Rotational transitions measured in this work are labeled with diamonds. The middle frame gives a closer look at the rotational transitions of the TiO isotopologues of J$^{\prime\prime}$=10 in the frequency range of 344--352\,GHz with calculated line strength (green) and experimental values for the intensity per laser shot (black) with respect to $^{48}$TiO. The upper frames show examples of measured line profiles (dots) and fitted Voigt profiles (blue solid line). \label{fig:over}}
	\end{center}
\end{figure*}

To each of the measured transitions a Voigt profile was fitted and  line-center frequencies were determined with $1\sigma-$uncertainties of less than 0.1\,MHz. 
The here presented molecular parameters of TiO were determined by a combined analysis of high-resolution data including electronic transitions in the optical range as well as rotational transitions in the GHz region. 
From the electronic spectra of  $^{48}$TiO we included nearly 8000 transitions of the $(A^3\Phi-X^3\Delta)$ system observed by Ram\etal\cite{Ram1999} and roughly 1300 transitions of the $(B^3\Pi-X^3\Delta)$ system investigated by Amiot\etal\cite{Amiot1995}.\added[id=A]{The measured data set were kindly provided by C. Amiot upon request.} 
In the GHz range, the $^{48}$TiO data set from Namiki\etal\cite{Namiki1998} and recently published data of \deleted[id=A]{all stable} titanium isotopologues \added[id=A]{($^{46,47,49,50}$TiO)} from Lincowski\etal\cite{Lincowski2016} are taken into account in addition to our high-resolution data from the present work. 
\added[id=A]{To combine the experimental datasets from different apparatus, a two-step fitting routine is used. In the first step the datasets are included with their reported frequency uncertainties as weighting factor \cite{LeRoy1998}. Subsequently, the sub-datasets are evaluated with respect to their weighted standard deviation value. This value was used to scale the weighting factor of the final fitting step. The literature data set errors are 1.021 (Amiot\etal\cite{Amiot1995}), 2.468 (Namiki\etal\cite{Namiki1998}), 1.234 (Ram\etal\cite{Ram1999}) and 1.209, 2.077, 2.478, 1.484 ($^{46}$TiO,$^{47}$TiO,$^{49}$TiO,$^{50}$TiO, Lincowski\etal\cite{Lincowski2016}).}
\deleted[id=A]{In the fitting procedure all transitions are weighted according to their line-position uncertainties \cite{LeRoy1998}.}
The program \texttt{PGOPHER} \cite{Western2016} is used to evaluate the $N^2$ Hamiltonian representation for all isotopologues.
Beside the pure rotational parameters ($B_{\nu}, D_{\nu}$, and $H_{\nu}$) for the ro-vibrational parametrization of the $X^3\Delta$ and $A^3\Phi$ state further parameters have to be considered such as the state origin ($T_{\nu}$), the spin-orbit coupling parameters ($A_{\nu}, A_{D_\nu}$),\added[id=A]{ the spin-rotation coupling parameter ($\gamma$),} and the spin-spin coupling parameter ($\lambda_{\nu},\lambda_{D_{\nu}}$). In addition, 
for the stable isotopes with a non-vanishing nuclear spin moment the hyperfine coupling parameters ($a_{\nu},b_{\nu},c_{\nu},b_{D,\nu},c_{D,\nu},eQq_{0_{\nu}}$)  have to be considered.  
It is known \cite{Lincowski2016} that the hyperfine spin-orbit interaction for the $^3\Delta_2$ state is perturbed by the a$^1\Delta$ state, which is represented in the fit by the parameter $\Delta a$. 
In the case of the $B^3\Pi$ state \deleted[id=A]{the spin-rotation coupling parameter ($\gamma$) and} the $\lambda$-doubling parameters ($o,o_D,o_H,p,p_D,q,q_D$) need to be taken into account. 
Considering the well-known mass dependencies for diatomic molecules shown by Dunham \cite{Dunham1932} and in related work by Ross and Watson\ \cite{Watson1973,Ross1974,Watson1980}, the size of the parameter space is reduced from nearly 200 individual molecular parameters to 69 mass-independent parameters. The following general equation was used in the Dunham-type multi-isotopologue analysis to constrain the molecular parameters \added[id=A]{(see also Breier\etal\cite{Breier2018})}

%\makeatletter 
%\def\@eqnnum{{\normalsize \normalcolor (\theequation)}} 
%\makeatother

{%\scriptsize 
\begin{eqnarray}
\operatorname{X}_{\nu,\alpha}=\sum_{k}\eta\cdot\mu_\alpha^{\,-\frac{2l+k}{2}}\cdot\operatorname{\hat{O}}_{k,l}\cdot\left(1+\sum_{i={A,B}}\frac{m_e}{M^i_\alpha}\Delta_{\operatorname{\hat{O}}_{k,l}}^i\right)_{BO}\cdot\left(\nu+\frac{1}{2}\right)^k.%\nonumber
\label{eq:Dun}
\end{eqnarray}}
The molecular parameter $\operatorname{X}_{\nu,\alpha}$ of isotopologue $\alpha$ in its vibrational state $\nu$ is represented by a sum over Dunham-like expansion terms.  Here $\operatorname{X}_{\nu,\alpha}$ is one of the parameters 
$T_{\nu,\alpha}$, $B_{\nu,\alpha}$, $D_{\nu,\alpha}$, $...$, given in Tab.~\ref{tab:operator}.
The index $k$ denotes the ro-vibrational coupling order and $l$ stands for the molecular parameter expansion order (e.g. $l$=0 $\rightarrow$ $U_{k,0}$  $\rightarrow$ $T_{\nu}$, and $l$=1  $\rightarrow$ $U_{k,1}$  $\rightarrow$ $B_{\nu}$). 
Each term in Eq.~\ref{eq:Dun} consists of five factors, the first being the effective proportional factor $\eta$.
For most parameters, this scaling factor is simply one, but for the hyperfine parameters $a$, $b$, $c$, and $c_I$ this factor is equal to the nuclear $g$ factor ($g_N$) and for the case of the $eQq_0$ parameter it is equal to the nuclear quadrupole moment $Q$, see \cite{Muller2015,Lattanzi2015} for more details.
For $^{47}$Ti and $^{49}$Ti, the used g$_n$ factors are  0.315392(4) and 0.315477(3) \cite{Stone2005,Harris2009}, respectively,  and are determined by their nuclear magnetic dipole moment.  Pyykk\"{o} \cite{Pyykko2008} reviewed the values for the nuclear quadrupole moments as Q($^{47}$Ti) =0.302(10)\,barn and Q($^{49}$Ti)= 0.247(11)\,barn. 
The second factor from Eq.~\ref{eq:Dun} represents the reduced mass-scaling factor $\mu_\alpha$ of an isotopologue $\alpha$ of the diatomic molecule \textit{AB}. Thereon the isotopic invariant factor $\operatorname{\hat{O}}_{k,l}$ follows, which is a placeholder for the actual molecular parameter to be fitted, e.g., in the case of  $\operatorname{X}_{\nu,\alpha}$ being $T_{\nu,\alpha}$ the invariant factor $\operatorname{\hat{O}}_{k,l}$ is $U_{k,0}$, see Tab.~\ref{tab:operator}. \added[id=A]{The subindexes of $\hat{O}_{kl}$ are mostly identical to former publications despite parameters reflecting the spin-rotational contributions like, $\gamma$ or $c_I$. These are shifted in the $l$-index following directly their reduced mass-dependency of $\mu^{-1}$\cite{Brown1977}.}
The fourth factor (term in \replaced[id=A]{parentheses}{brackets}) \replaced[id=A]{handles }{is} the Born-Oppenheimer breakdown (BO) \deleted[id=A]{term} which is different for each isotopic invariant parameter. The sum index $i$ refers to the atoms A and B of the diatomic molecule, e.g., A=Ti and B=O. Further, $m_e$ is the electron mass, $M^A$ is the mass of atom A and $\Delta_{\operatorname{\hat{O}}_{k,l}}^A$ is the to-be-fitted BO coefficient of atom A. The same procedure applies to atom B. Our dataset leads to the BO terms of the first-order rotational expansion term $U_{01}$ and the centrifugal-distortion term of the spin-orbit coupling $A_{01}$ with respect to titanium. The last factor of Eq.\ref{eq:Dun} describes the ro-vibrational coupling of the expansion terms.
As an example $X_{\nu,\alpha}$ for $\nu$=0,  $\alpha$=$^{48}$Ti$^{16}$O and $l$=1, i.e. $B_{0,^{48}Ti^{16}O }$, is 
\begin{subequations}
\begin{eqnarray}  
	B_{0}  &=&    \mu^{-1}U_{01}\cdot\left(1+\frac{m_e}{M^{Ti}}\cdot\Delta_{U01}^{Ti}+\frac{m_e}{M^{O}}\cdot\Delta_{U01}^{O}\right)_{BO}+  \label{eq:Dun2a} \\
	&&\mu^{-1.5}U_{11}\cdot\left(0+\frac{1}{2}\right)+\mu^{-2}U_{21}\cdot\left(0+\frac{1}{2}\right)^2+\mu^{-2.5}U_{31}\cdot\left(0+\frac{1}{2}\right)^3   \label{eq:Dun2b} \\
	& \equiv & Y_{01}  +  Y_{11}  \cdot\left(0+\frac{1}{2}\right) +  Y_{21}   \cdot\left(0+\frac{1}{2}\right)^2 +  Y_{31}   \cdot\left(0+\frac{1}{2}\right)^3   \label{eq:Dun2c} \\
 \text{with}  \nonumber  \\
\mu  &=& \frac{M^{Ti} M^{O}}{M^{Ti}+M^{O}} \text{\hspace{1cm}(see Tab.~\ref{tab:mass_re})}. \nonumber
\end{eqnarray} \label{eq:Dun2}
\end{subequations}
Note that not only the mass-independent $U_{ij}$ coefficients are given here but also the mass-dependent Dunham coefficients $Y_{ij}$ are shown  in Eq.~\ref{eq:Dun2c} and they will be discussed in Sec.~\ref{ssec:struc}. The BO contributions in the terms~\ref{eq:Dun2b} are set to \replaced[id=A]{zero}{unity}. 
 
In our analysis we have used mass units \replaced[id=A]{that have been compiled and referenced in the latest evaluation of atomic masses, AME2016}{as given by} \cite{Wang2017}\deleted[id=A]{ from the year 2003}.\deleted[id=A]{In the course of our measurements and data analysis a new set of recommended mass values became available%  \cite{Wang2012}
	 which differs by only $10^{-8}$ from the previous data set. For consistency reasons with former publications% \cite{Brunken2008,Kania2011}
we decided to use the 2003 mass values in our present analysis. The accuracy of the derived mass-independent Dunham parameters ($\hat{O}_{k,l}$) is not affected by this.} 
\added[id=A]{One should notice, that for a single-isotope fit, the effects on A$_D$ and $\gamma$ are not distinguishable in terms of energy in a $^3\Delta$ electronic state. Performing an isotopic invariant fitting procedure allows to distinguish between these contributions \cite{Muller2015}. Therefore, two different fitting procedures were used: \textit{Fit A} is without the contribution of the spin-rotational parameter $\gamma$  while  \textit{Fit B} is taking the contribution of $\gamma$ into account (see Tab.\ref{tab:parameter}). For the first time a spin-rotational value for TiO is determined as 233.8(17)\,MHz. However, in the following discussion section, we stick to the mass-independent parameter set of \textit{Fit A}, which resulted in a smaller value of the weighted error. Furthermore, comparing our results to former published data \cite{Ram1999,Lincowski2016} is straight forward when using \textit{Fit A}.}
\added[id=A]{The results of our \texttt{PGOPHER} simulations are available as supplementary material.}
\begin{table}
	\caption{Translation of the \replaced[id]{effective}{commonly used} molecular parameters $\operatorname{X}_{\nu,\alpha}$ to the mass-independent Dunham-like parameters $\operatorname{\hat{O}}_{k,l}$ and the effective proportional factor $\eta$ as given in Eq.~\ref{eq:Dun}.\label{tab:operator}}
	\begin{center}	 
		\begin{threeparttable}
			\begin{tabular}{*{3}{l}|*{3}{l}}
				\toprule
				\multicolumn{1}{l}{$\operatorname{X}_{\nu,\alpha}$}&\multicolumn{1}{l}{$\operatorname{\hat{O}}_{k,l}$}&\multicolumn{1}{l}{$\eta$}&\multicolumn{1}{l}{$\operatorname{X}_{\nu,\alpha}$}&\multicolumn{1}{l}{$\operatorname{\hat{O}}_{k,l}$}&\multicolumn{1}{l}{$\eta$}\\%\hline
				\midrule 
				%&&&&&&&\\
				$T_{\nu,\alpha}$&$U_{k,0}$&1&$\lambda_{D_{\nu,\alpha}}$&$\lambda_{k,1}$&1\\
				$B_{\nu,\alpha}$&$U_{k,1}$&1&$a_{\nu,\alpha}$&$a_{k,0}$&g$_N$\\
				$D_{\nu,\alpha}$&$U_{k,2}$&1&$\Delta a_{\nu,\alpha}$&$\Delta a_{k,0}$&g$_N$\\
				$H_{\nu,\alpha}$&$U_{k,3}$&1&$b_{\nu,\alpha}$&$b_{k,0}$&g$_N$\\
				$\gamma_{\nu,\alpha}$&$\gamma_{k,1}$&1&$c_{\nu,\alpha}$&$b_{k,0}$&g$_N$\\
				$A_{\nu,\alpha}$&$A_{k,0}$&1&$b_{D_{\nu,\alpha}}$&$b_{k,1}$&g$_N$\\
				$A_{D_{\nu,\alpha}}$&$A_{k,1}$&1&$c_{D_{\nu,\alpha}}$&$c_{k,1}$&g$_N$\\
				$\lambda_{\nu,\alpha}$&$\lambda_{k,0}$&1&$eQq_{0_{\nu,\alpha}}$&$eQq_{0_{k,0}}$&Q\\
				\bottomrule
			\end{tabular}
		\end{threeparttable}
	\end{center}
\end{table}

\section{Quantum-chemical calculations}

High-accuracy quantum-chemical calculations were carried out to support the experimental investigations. All of them were performed at the coupled-cluster (CC) level \cite{Shavitt2009} using large correlation-consistent basis sets \cite{Dunning1989}. To be more specific, the equilibrium distance of TiO was determined using a composite scheme together with basis-set extrapolation, as described in Refs. \cite{Heckert2005,Heckert2006}. The extrapolation at the Hartree-Fock level \cite{Feller1993} was based on results cc-pwCVXZ (X=T,Q,and 5) basis sets \cite{Peterson2002,Balabanov2005}, while the extrapolation at the CCSD(T) level \cite{Raghavachari1989,Helgaker1997} has been carried out using the cc-pwCQZ and cc-pwC5Z sets. Additional corrections account for the 
difference of CCSD(T) to the full CC singles, doubles, triples (CCSDT) treatment \cite{Watts1990} (as computed using a cc-pVTZ basis) and for the effect of quadruple excitations at the CC singles, doubles, triples, quadruples (CCSDTQ) level \cite{Kallay2001} (as computed at the cc-pVDZ level). Furthermore, scalar-relativistic corrections were considered at the spin-free X2C-1e level \cite{Dyall1997,Kutzelnigg2005,Ilias2007,Cheng2011}; these calculations were carried out with an uncontracted ANO-RCC basis set \cite{Roos2004,Roos2005}.

Based on previous experience \cite{Heckert2006}, the overall accuracy of the computed TiO bond distance should be in the range of about 0.002 \AA. This accuracy is for most spectroscopic purposes sufficient, but it does not match the accuracy provided by a Dunham analysis. Thus, we refrain from a detailed computational study of the other spectroscopic parameters, though in principle possible, and in the following only provide computational data for the Born-Oppenheimer breakdown or correction parameters. These parameters were computed as outlined in Refs. \cite{Gauss2010,Puzzarini2013}. The required rotational g-tensor \cite{Gauss2007} were determined at the frozen-core  CCSD(T) level using the aug-cc-pVTZ basis \cite{Woon1993}, while the adiabatic correction to the bond distance is obtained via the computation of the diagonal Born-Oppenheimer correction \cite{Gauss2006} (performed at the CC singles, doubles (CCSD) level using the cc-pwCVTZ basis \cite{Dunning1989,Balabanov2005}). The usually rather small Dunham correction was not considered in the present work.

All calculations were carried out using the CFOUR program package\,\cite{CFOUR}.

\section{Discussion and Conclusion}
In this section the discussion is based on experimental measurements and on quantum-chemical calculations. In Sec.\,\ref{ssec:dunham} we apply the mass-independent Dunham-like parametrization to TiO. Section~\ref{ssec:struc} is concerned with the analysis of the TiO molecular structure dominated by the mass-independent rotational parameter $U_{01}$ (see Eq.~\ref{eq:Dun2}) and the corresponding BO terms. The topic of  Sec.~\ref{ssec:hyper} is the electronic structure of the TiO molecule. Based on the free atom comparison method, the molecular bonding is investigated by analysis of the molecular hyperfine parameters of the odd TiO isotopologues.
\begin{table}
	\caption{Comparison of mass-depending molecular parameters and bond lengths for the various TiO isotopologues.\label{tab:mass_re}}	
	%\begin{adjustbox}{totalheight=\textheight-2\baselineskip, max width=0.99\textwidth}
	\begin{threeparttable}
		\begin{tabular}{*{1}{l}*{1}{D{.}{.}{2.13}}*{1}{D{.}{.}{2.8}}*{1}{D{.}{.}{1.12}}*{1}{D{.}{.}{1.13}}}
			\toprule
			&\multicolumn{1}{c}{$\mu_\alpha$(u)}&\multicolumn{1}{c}{$\tilde{A}_{01}$($10^5$cm$^{-1}$)}&\multicolumn{1}{c}{Y$_{01}$(cm$^{-1}$)}&\multicolumn{1}{c}{r$_e$(\AA)}\\%\hline
			\midrule 
			%&&&&&&&\\
			$^{46}$Ti$^{16}$O&11.865012325(12)&-2.6763(27)&0.541147722(47)&1.620340280(71)\\
			$^{47}$Ti$^{16}$O&11.930564845(8)&-2.6370(27)&0.538175512(46)&1.620338581(70)\\
			$^{48}$Ti$^{16}$O&11.993884458(8)&-2.5997(26)&0.535335382(46)&1.620336957(70)\\
			$^{49}$Ti$^{16}$O&12.055488302(7)&-2.5640(26)&0.532600833(46)&1.620335394(70)\\
			$^{50}$Ti$^{16}$O&12.115047227(8)&-2.5301(26)&0.529983487(46)&1.620333898(70)\\
			$^{48}$Ti$^{18}$O&13.086589627(9)&-4.0629(130)&0.490647357(47)&1.620318041(77)\\
			\bottomrule
		\end{tabular}
	\end{threeparttable}
	%\end{adjustbox}
\end{table}

\subsection{Mass-independent Duham-like parameterization}\label{ssec:dunham}
High-resolution measurements of stable isotopolgues allow the determination of ground-state molecular parameters to high precision, see Tab. \ref{tab:mass_re}. The value of the first-order expansion term of the rotational constant for the main isotopologue (i.e., the $U_{01}$ containing term of  $B_0$ see Eq.~\ref{eq:Dun2a}) is determined as 0.535335379(46)\,cm$^{-1}$ and agrees well with the value observed by Ram\etal of 0.53533536(16)\,cm$^{-1}$ (see $p_0$ value of $B_{\nu}$ in Tab.~7 of \cite{Ram1999}).
Thanks to the available large data set for the main isotopologue, and applying mass-invariant scaling, the same high accuracy can be reached 
for the parameters of the rare isotopologues. 
The consistency of the determined  mass-invariant parameters can be checked by applying the so-called Kratzer~\cite{Kratzer1920} and Pekeris~\cite{Pekeris1934} relation: $U_{02}=4U_{01}^3/U_{10}^2$, which results in 8.676010(16)$\cdot10^{-5}$\,cm$^{-1}$\,u$^2$.
The direct determination of $U_{02}$ using Eq.~{\ref {eq:Dun}} yields  8.67181(37)$\cdot10^{-5}$\,cm$^{-1}$\,u$^2$ and thus is in good agreement with the above value given that the BO correction terms are not considered in this approach. 
Furthermore, we can easily extract the dissociation-energy value of 6.92053(56)\,eV for the ground state from the relation $\mathcal{D}_e = \frac{\omega_e^2}{4\omega_e \chi_e} \, \hat{=} \,  \frac{U_{10}^2}{4 U_{20}}$  \text{(\cite{Gordy1984})}. Despite the restriction to a simple Morse potential \cite{Morse1929}, the result is in very good agreement with the values derived from crossed-beam studies by Naulin\etal\cite{Naulin1997a} with  $\mathcal{D}_e = $ 6.87$^{+0.07}_{-0.05}$\,eV and as well from a potential analysis of the electronic ground-state performed by Reddy\etal\cite{Reddy2000} resulting in 6.94(16)\,eV. 
 \begin{table}[b]
 	\caption{Comparison of the different $r_e$\tnote{a} distances for TiO\label{tab:re}}	
 \begin{center} 
 	\begin{threeparttable}
 		\begin{tabular}{*{1}{l}*{1}{D{.}{.}{1.13}}*{1}{l}}
 			\toprule 
 			&\multicolumn{1}{c}{bond length (\AA)}&\multicolumn{1}{c}{(Comment)}\\%\hline
 			\midrule 
 			%&&&&&&&\\
 			$r_e^{48}$&1.62033709(25)&Ram \textit{et al.}\cite{Ram1999}\\
 			$r_e^{48}$&1.620336957(70)&This work\\
 			$\overline{r}_e$&1.62033386(527)&This work\\
 			$r_e^{\scalebox{0.5}{BO}}$&1.62009060(31)&This work\\
 			$r_{e,theo}^{\scalebox{0.5}{BO}}$&1.6189&This work\\
 			\bottomrule
 		\end{tabular}
 	\end{threeparttable}
 \end{center}
 \end{table}
\subsection{Molecular structure of ($X^3\Delta$) TiO}\label{ssec:struc}
From measurements of six TiO isotopologues the bond length of TiO can be obtained via the moment of inertia $I_e=\mu r_e^2$ and the related rotational constant $B_e$ ~\cite{Gordy1984}.
In the literature $B_e$ is derived from the mass-dependent Dunham coefficient $Y_{01}$, provided that $Y_{01}\approx B_e$ is a good approximation \cite{Dunham1932}. The $r_e$ bond length of six TiO isotopologues are given in Tab.~\ref{tab:mass_re} and they significantly scatter about a mean value $\overline{r}_e$ of 1.62033386(527)\,\AA. For $^{48}$Ti$^{16}$O Ram\etal published a bond length $r_e^{48} = 1.62033709(25)$\,\AA\ which is in excellent agreement with the value derived from the present study. The derivation of isotopic-specific bond lengths are clear indications for deviations from the Born-Oppenheimer approximation. A mass-independent approach which includes the Born-Oppenheimer breakdown (BO) coefficients (see Eq.~\ref{eq:Dun2a}) is given by
\begin{eqnarray}
	B_e = U_{01}\cdot \mu^{-1} \qquad \text{and}\qquad r_e^{\scalebox{0.5}{BO}}=U_{01}^{-\nicefrac{1}{2}}\cdot\left(\frac{\hbar}{4\pi}\right)^{\nicefrac{1}{2}},
\end{eqnarray}
where  r$_e^{\scalebox{0.5}{BO}}$ is the mass-independent BO corrected bond length, which is given as  $r_e^{\scalebox{0.5}{BO}}=4.1058043277(10)U_{01}^{-\nicefrac{1}{2}}$
with $U_{01}$ given in units of atomic mass units $u$ and wavenumbers cm$^{-1}$, and the distance given in \AA.

The derived bond length is 1.62009060(31)$~\AA$ and shown in Tab.\,\ref{tab:re}.  \replaced[id=A]{The}{Interestingly, the} BO distance is significantly smaller (by about 0.00024\,\AA) than all the mass-dependent values\added[id=A]{, as is usually the case}.\\ 
This value should be also compared to the best theoretical estimate for the TiO bond distance of 1.6189\,\AA\ as obtained using a composite approach with corrections for scalar-relativistic effects. The theoretical value thus is too short in comparison with experimentally derived distance, but the discrepancy is within the usual error margin of high-accuracy state-of-the-art predictions \cite{Puzzarini2010}. It is also interesting to note that without consideration of scalar-relativistic effects an even shorter distance is obtained, namely 1.6176\,\AA, scalar-relativistic effects thus amount to about 0.0012\,\AA\ and are not negligible. Furthermore, the importance of higher excitations should be noted, as pure CCSD(T) computations (extrapolated to the basis-set limit) yield 1.6099\,\AA, i.e., a value which is about 0.008\,\AA\ too short compared to the best non-relativistic value. Corrections for a full triple excitations treatment amount to 0.004\,\AA, the corrections for quadruple excitations are of the same order of magnitude. 

Titanium is a group 4 element in the periodic table, such as zirconium (Zr) and hafnium (Hf). By comparing  the $r_e$ value of $^{48}$TiO to those of the most abundant isotopologues of ZrO and HfO (see Tab. \ref{tab:dipole_re}) it is seen that the bond length increases with the row number. 
For all three oxides, the determined mass-independent $r_e^{\scalebox{0.5}{BO}}$ values are smaller than the respective mass-dependent $r_e$ values,  
which is a consequence of the electron-nucleus interaction that is considered by the BO breakdown term. 
Note that the difference $|r_e - r_e^{\scalebox{0.5}{BO}}|$  is of same order for TiO,  ZrO, and HfO.\\
Looking at monoxides of the neighboring elements in the periodic table. i.e.,  scandium monoxide (ScO) and vanadium monoxide (VO), it is noted that the bond lengths decrease with increasing atomic number, 
which can be attributed to the influence of $\delta$-orbital\,\cite{Bridgeman2000}. 
 \begin{table*}[h]
 	\caption{Comparison of dipole moment, $r_e$\tnote{a} and mass independent Born-Oppenheimer correction term for $^{48}$Ti$^{16}$O and related main isotopologues of transition metal diatomic molecules.\label{tab:dipole_re}}
 	\begin{center}	 
 	\begin{threeparttable}
 		\begin{tabular}{*{1}{l}*{1}{D{.}{.}{1.7}}*{1}{D{.}{.}{1.12}}*{1}{D{.}{.}{1.12}}*{2}{D{.}{.}{2.7}}}
 			\toprule
 			{Molecule (AB)}&\multicolumn{1}{c}{$\mu$(D)}&\multicolumn{1}{c}{r$_e$(\AA)}&\multicolumn{1}{c}{r$_e^{\scalebox{0.5}{BO}}$(\AA)}&\multicolumn{1}{c}{$\Delta_{U01}^\textnormal{A}$}&\multicolumn{1}{c}{$\Delta_{U01}^\textnormal{B}$}\\%\hline
 			\midrule 
 			%&&&&&&&\\
 			TiO(X$^3\Delta$)\tnote{a}&3.34(1)&1.62033696(7)&1.62009060(31)&-8.253(24)&-6.112(8)\\
 			ZrO(X$^1\Sigma^+$)\tnote{b}&2.551(11)&1.71195242(73)&1.711 745 27(74)&-4.872(39)&-6.189(3)\\
 			HfO(X$^1\Sigma^+$)\tnote{c}&3.431(5)&1.7231481&1.7229734&-3.40(57)&-5.656(23)\\
 			ScO(X$^2\Sigma^+$)\tnote{d}&4.55(8)&1.66608(19)&\multicolumn{1}{c}{-}&\multicolumn{1}{c}{-}&\multicolumn{1}{c}{-}\\
 			VO(X$^4\Sigma^-_{1/2}$)\tnote{e}&3.355(14)&1.5893(2)&\multicolumn{1}{c}{-}&\multicolumn{1}{c}{-}&\multicolumn{1}{c}{-}\\
 			\bottomrule
 		\end{tabular}
 		\begin{tablenotes}
 			\footnotesize
 			\item [a] This work, but $\mu$ value taken from Steimle\etal\cite{Steimle2003}\\
 			\item [b] Values taken from Beaton\etal\cite{Beaton1999}, Lesarri\etal\cite{Lesarri2002}, and Suenram\etal\cite{Suenram1990}\\
 			\item [c] Values taken from Lesarri\etal\cite{Lesarri2002} and Suenram\etal\cite{Suenram1990}\\
 			\item [d] Values taken from Childs\etal\cite{Childs1988}, Shirley\etal\cite{Shirley1990}, and Mukund\etal\cite{Mukund2012}\\
 			\item [e] Values taken from Flory\etal\cite{Flory2008}, Suenram\etal\cite{Suenram1991}, and Huber\etal\cite{Huber1979}\\
 		\end{tablenotes}
 	\end{threeparttable}
 	\end{center}
 \end{table*}

By taking a look at the rotational BO correction term $\Delta_{U01}^\textnormal{A}$ (A = Ti, Zr, Hf) it can be seen (Tab.~\ref{tab:dipole_re}) that for titanium the value -8.253(24) is in absolute terms larger than those of the elements zirconium and hafnium being -4.872(39) and -3.40(57), respectively \cite{Lesarri2002}, i.e., with increasing row number, the rotational BO correction increases. 
The rotational BO correction term $\Delta_{U01}^\textnormal{B}$ (B=oxygen) is similar to the values reported for ZrO and HfO. 
The magnitude of the BO correction term, see $(...)_{BO}$ term in Eq.~\ref{eq:Dun2a},  scales with the inverse of the atomic mass, and its effect on the rotational constant $B_0$ for TiO is roughly 0.1\,\permil . The influence of the BO correction term on $B_0$ for TiO is ten times larger than for HfO.\\
An even larger effect of isotopic scaling shows the BO correction on the distortion term of the spin-orbit coupling $A_{01}$. For TiO the BO term increases the $A_{01}$ value by around 0.6\,\%. 
Unfortunately, no $A_{01}$-BO corrected values for other diatomic molecules with spin-orbit coupling are known and therefore a comparison with TiO is not possible.  
\\
The above mentioned Born-Oppenheimer breakdown corrections are crucial when predicting rotational constants (Tab.~\ref{tab:measurment_44}) and transition frequencies (Tab.~\ref{tab:prediction_44}) of isotopologues that are not easy to access via laboratory investigations. For example, astronomical searches for the short-lived $^{44}$TiO isotopologue, expected to be present in supernovae remnants, require an  accuracy in the sub-MHz range which is only feasible with consideration of the BO correction terms. 

Finally, we compare the experimentally derived BO correction terms with those from theory. There, the corresponding values are $\Delta_{U01}^{Ti}= -4.9$ and $\Delta_{U01}^O= -6.7$. Thus, a rather good agreement is seen for the breakdown parameter of oxygen, while the experimentally determined parameter for titanium is significantly larger than the computed one. However, it should be noted that the uncertainty in the theoretical value for titanium is rather high, e.g., using rotational g-tensors from CCSD instead of CCSD(T) computations changes the titanium parameter by roughly -0.8. Furthermore, the theoretical approach to the BO breakdown parameters does not consider spin-orbit effects and thus might be insufficient considering the large BO effect on the $A_{01}$ term.
\begin{table}
	\caption{Predicted rotational parameter (in MHz) of the X$^{3}\Delta$ state of $^{44}$TiO\tnote{a}.\label{tab:measurment_44}}	 
	\begin{threeparttable}
		\begin{tabular}{*{1}{l}*{1}{D{.}{.}{2.14}}}
			\toprule
			&\multicolumn{1}{c}{$^{44}$TiO}\\%\hline
			\midrule 
			%&&&&&&&\\
			$T \cdot 10^{-7}$&1.5262874(8)\\
			$B \cdot 10^{-4}$&1.636608750(37)\\
			$D \cdot 10^2$&1.895723(78)\\
			$H \cdot 10^9$&3.60(32)\\
			$A \cdot 10^{-6}$&1.5184807(15)\\
			$A_D \cdot 10^{1}$&-8.4373(42)\\
			$\lambda\cdot 10^{-4}$&5.23787(19)\\
			$\lambda_D \cdot 10^2$&1.697(38)\\
			\bottomrule
		\end{tabular}
		\begin{tablenotes}
			\footnotesize
			\item [a] Values in brackets represents $1\sigma$ uncertainties.
		\end{tablenotes}
	\end{threeparttable}
\end{table}
\\

\subsection{Electronic structure of ($X^3\Delta$) TiO}\label{ssec:hyper}
From the analysis of $^{47}$TiO and  $^{49}$TiO the hyperfine (hf) parameters were derived, as given in Tab.~\ref{tab:measurment_hyper}. \added[id=A]{The hf values of the Frosch and Foley \cite{Frosch1952}, i.e., the hyperfine constant b, as well as, the dipolar interaction constant c differ from those reported by Linkwoski\etal\cite{Lincowski2016}. However, we re-fitted the molecular parameters of the odd numbered TiO isotopologues with \texttt{PGOPHER} only by using their rotational transitions, which are in good agreement with the values derived from the mass-independent fitting routine, as can be seen in Tab.~\ref{tab:measurment_hyper}.}\\
Anyway, the molecular hf parameters allow insights into the electronic structure of TiO. Four topics will be discussed in this section: The ionic vs. atomic character of the TiO bond, the contribution of the various atomic orbitals to the molecular bond, the dipole character of the molecular bond, and the localization of the bonding electrons. 
For further analysis of the electronic structure we apply the free atom comparison method (FACM)\cite{Fletcher1993,Namiki2002,Namiki2003,Namiki2004} which makes use of the most dominant electronic configuration and its orbital model (details are given in the \hyperref[sec:app]{Appendix}).\added[id=A]{For the first time, this analysis of the molecular structure of TiO is based only on experimental values.} 
\\
%%%%%%%%%%%%%%%%%%%%%%
{\bf Background}. The electron configuration of the ground state $X^3\Delta$ can be qualitatively described as $(\textrm{core})(9\sigma)^1(1\delta)^1$,\replaced[id=A]{ the open-shell electrons occuping the $9\sigma$ and the $1\delta$  molecular orbital}{with $(\textrm{core}) =(1\sigma)^2-(8\sigma)^2(1\pi)^4-(3\pi)^4$}. Bauschlicher\etal\,\cite{Bauschlicher1983}concluded in their theoretical study that the two unpaired electrons originate from the titanium. 
Namiki\etal\cite{Namiki2004} introduced a simple model for the unpaired electron orbitals\added[id=A]{, handling those molecular open-shell orbitals by describing them as a linear combination of titanium atomic/ionic orbitals.} 
\deleted[id=A]{In short, in their model one open-shell electron occupies the $9\sigma$  molecular orbital, which is described as a linear combination of titanium atomic orbitals, $\ket{9\sigma} = c_{4s}^{9\sigma}\ket{4s}-c_{3d}^{9\sigma}\ket{3d_\sigma}-c_{4p}^{9\sigma}\ket{4p_\sigma}$, with the normalization coefficients $c_{nx}^{9\sigma}$ related by $(c_{4s}^{9\sigma})^2+(c_{3d}^{9\sigma})^2+(c_{4p}^{9\sigma})^2=1$. The second unpaired electron of TiO, occupying the $1\delta$ molecular orbital, is basically the Ti $3d_\delta$ orbital, $\ket{1\delta} = \ket{3d_\delta}$.}
\\
%%%%%%%%%%%%%%%%

{\bf Ionic vs. atomic character of the bond}. 
Theoretical studies showed that the bonding of TiO is in between a covalent and an ionic one \cite{Bauschlicher1983,Bauschlicher1995}. \replaced[id=A]{The bond character is experimentally determined by using the FAC method together with the fine and hyperfine parameters of TiO, Ti, and Ti$^+$.}{Considering this, the orbitals are expressed as a linear combination of atomic and ionic orbitals $\ket{\chi}=c_{\textrm{atom}}\ket{\chi(\textrm{Ti})}+c_{\textrm{ion}}\ket{\chi(\textrm{Ti}^+)}$ with $\chi ={4s,4p_\sigma, 3d_\sigma ,3d_\delta}$. Here, the overlap integrals of atomic and ionic orbitals are neglected and the $\ket{\chi}$ functions are normalized  using $(c_{\textrm{atom}})^2+(c_{\textrm{ion}})^2=1$, see Namiki\etal\cite{Namiki2004}.} 
The nuclear spin-orbit interaction of a diatomic molecule can be expressed by  
\begin{equation}
a= 2 \mu_Bg_N\mu_N\frac{\mu_0}{4\pi}\frac{1}{\Lambda}\bra{\Lambda\Sigma}\sum_i\frac{\hat{l}_{zi}}{r_i^3}\ket{\Lambda\Sigma}
\end{equation} 
with the molecular nuclear spin-orbit constant $a$, the electron Bohr magneton $\mu_B$, the nuclear \textit{g} factor $g_N$, the nuclear Bohr magneton $\mu_N$,  the magnetic constant $\mu_0$, the $\Lambda$ orbital quantum number of the electronic state and the z component of the one-electron orbital angular operator $\hat{l}_{zi}$ of the electron in question. The expectation value from a $\sigma$-type orbital is zero, therefore only the $1\delta$ electron of TiO in the electronic ground state contributes to the nuclear spin-orbit interaction: $a = a(1\delta)$. 
The character of the $1\delta$ molecular orbital is defined by the titanium $3d$ orbital that, according to the FAC method, leads to an  ionic character, i.e., a $c_{ion}$ value above 0.7, more precisely, the  $c_\textrm{ion}$ coefficient 
can be calculated (see \ref{eq:a_delta}\,ff) via 
\begin{equation}
a(^{47}TiO)= (1-c_\textnormal{ion}^2)a_{3d}^{01}(^{47}Ti) + c_\textnormal{ion}^2a_{3d}^{01}(^{47}Ti^+).
\label{eq:nucspin}
\end{equation}
with $a_{3d}^{01}(X/X^+)$ being atomic/ionic nuclear spin-orbit coupling constant\deleted[id=A]{, see Tab.~\ref{tab:facm}}.  \deleted[id=A]{In the molecular frame the nuclear spin-orbit parameter $a$ is directly related to the $a_{3d}^{01}$ constants.}  
\deleted[id=A]{The $(3d^34s)$ electron configuration of Ti is used for the ground-state description of TiO, as proposed by Bauschlicher\etal~\cite{Bauschlicher1983}.
The atomic spin-orbit parameter $a_{3d}^{01}(Ti)$ in the $(3d^34s)$ electronic configurations is derived from scaling relations given by Aydin\etal~\cite{Aydin1989} based on their atomic titanium measurements.
From the work of Bouazza\cite{Bouazza2013} similar scaling relations for the ionic spin-orbit parameter $a_{3d}^{01}(Ti^+)$ are available for the titanium ion.}
The atomic~\cite{Aydin1989} and  ionic~\cite{Bouazza2013} parameters of titanium ($a_{3d}^{01}(Ti)$, $a_{3d}^{01}(Ti^+)$)in the specific electronic configurations (\hyperref[sec:app]{Appendix}) are derived from literature and are listed in Table \ref{tab:facm}.
From our experimentally derived nuclear spin-orbit parameter $a(^{47}TiO)=-53.155(43)$\,MHz, the ionic character of TiO is derived applying Eq.\ \ref{eq:nucspin} which yields $c_\textnormal{ion} = 0.799(151)$, a slightly more ionic character than the value of Namiki\etal\cite{Namiki2004}, $c_\textnormal{ion}=0.789(4)$. 
The large uncertainty of the derived $c_{ion}$ value is due to the uncertainty of the ionic nuclear spin-orbit parameter $a_{3d}^{01}(Ti^+)$, which was \added[id=A]{generally} not considered in the work of Namiki\etal\cite{Namiki2003}. 
Alternatively, the ionic character of the TiO bond can be derived from the molecular \added[id=A]{spin-orbit} fine-structure parameter A\deleted[id=A]{, which describes the coupling of the electron spin to the orbital angular momentum},
\replaced[id=A]{in a similar way as}{Similar to} the hyperfine parameter $a$ \added[id=A]{is derived} (see \ref{app:sin_orbit}\,ff) \replaced[id=A]{. By making use of the atomic and ionic fine structure parameters $\zeta_{\textrm{3d}}$ of titanium  (see Tab. \ref{tab:facm}),}{ the fine-structure parameter $A$ is correlated to the atomic and ionic parameters $\zeta_{\textrm{3d}}$,} 
we obtain an ionic character coefficient $c_\textnormal{ion} = 0.813(56)$, which is slightly larger than the value obtained from $a(^{47}TiO)$. 
In the proceeding analysis, we use the more accurate value derived from $A(^{47}TiO)$. Furthermore, we used the $c_\textnormal{ion}$ parameter from our analysis, and atomic and ionic parameters from the literature to derive scaled molecular parameters, as given in Tab.~\ref{tab:facm}.
\\
%%%%%%%%%%%%%%%%%%

{\bf Contribution of the atomic orbitals to the molecular bonding}. 
A close look at the molecular Fermi-contact parameter $b_\textnormal{F}(=b +\nicefrac{c}{3})$ \deleted[id=A]{via the relation $b_\textnormal{F}(TiO)= \frac{1}{2}(b_\textnormal{F}(9\sigma)+b_\textnormal{F}(1\delta))$} reveals that the contribution from the $1\delta$ electron is small compared to the $9\sigma$ orbital \replaced[id=A]{$(39.5(37)\textnormal{\,MHz}>-718(33)\textnormal{\,MHz})$}{, i.e., $b_\textnormal{F}(9\sigma) > b_\textnormal{F}(1\delta)=a_{\textrm{3d}}^{10}(TiO)=39.5(37)\textnormal{\,MHz}$} (see \hyperref[sec:app]{Appendix}). 
The $\sigma$ character $c_{4s}^{9\sigma}$ of the $9\sigma$ orbital is evaluated by assuming that the non $s$-type atomic orbitals ($p_\sigma, d_\sigma$) are compensating each other (i.e., last two terms in \ref{app:bF_sigma} and \ref{app:bF} cancel out each other)\replaced[id=A]{,}{.} 
and leads to the value of $c_{4s}^{9\sigma}= 0.868(20)$.
This agrees well with the pronounced $4s$ orbital character of Namiki\etal\cite{Namiki2004} expressed by $c_{4s}^{9\sigma}=0.895(1)$. 
The comparison of those two values shows that if we assume `no-other $\sigma$-type' contribution to $c_{4s}^{9\sigma}$ we slightly underestimate the $\sigma$ character of the $4s$ orbital. It can also be seen that our value has a large uncertainty compared to the one by Namiki\,\textit{et al.}\added[id=A]{, referring to their neglection of experimental atomic/ionic uncertainties}. 
The missing contributions to the $9\sigma$ molecular orbital from the non $s$-type atomic orbitals, i.e., $p_\sigma$ and $d_\sigma$, 
are assumed to be equal, so that $\left({c_{3d_\sigma}^{9\sigma}}\right)^2=\left({c_{4p_\sigma}^{9\sigma}}\right)^2=0.12(2)$.
Furthermore, it follows, that the mixed Fermi-contact $4p$ parameter value is equal to the $3d$ one, but of opposite sign, i.e., $a_{3d}^{10}(\textnormal{TiO}) = - a_{4p}^{10}(\textnormal{TiO})$, see \ref{app:bF_sigma}.  
\\
%%%%%%%%%%%%%%%%%%%

{\bf The dipole character of the molecular bond}. 
The \added[id=A]{hyperfine} molecular dipole \replaced[id=A]{interaction}{Fermi-contact} parameter $c$ can be used to investigate the dipole character of the molecular bond\deleted[id=A]{. This can be done using the FAC method via $c(TiO)=\frac{1}{2}(c(9\sigma)+c(1\delta))$ which can be expressed as  $c = \mathcal{T}(a_{3d}^{12}(\textnormal{TiO})) + \mathcal{T}(a_{4p}^{12}(\textnormal{TiO})) $ by separating the 3d and 4p orbital contributions of $c(9\sigma)$ and $c(1\delta)$ in two terms}, see \ref{app:c}\,ff and specifically \ref{app:2c}.  From this it can be seen that the dipole character  predominantly results from the $3d$ orbital term with 16.8(14)\,MHz being about 1.5 times larger than the $4p$ orbital term with 11.6(14)\,MHz. The mixed dipolar parameter for the $4p$ orbital $a_{\textrm{4p}}^{12}(TiO)$ is then determined to be $157(29)$\,MHz. Hence, the contribution of the 4p orbitals to the $9\sigma$ MO is not negligible.\\
In summary, it seems that the 9$\sigma$ molecular orbital is pushed towards the titanium electrostatically due to the anionic character of the oxygen and mixes with the $4p$ and $3d$ atomic orbital of Ti.      
This can be described as a polarization effect that leads to an increased molecular dipole parameter $c$.
\\
%%%%%%%%%%%%%%%%%%%%

{\bf The localization of the bonding electrons between Ti and O}.
The electric quadrupole parameter  $eQq_0$ originates from the electric-field gradient from all electrons within the molecule on the odd-numbered Ti nuclei. 
It can be thought to be build of two parts $ eQq_0=eQq_{0\textnormal{|pol}}+eQq_{0\textnormal{|el}}$, with  $eQq_{0\textnormal{|pol}}$ as the static polarization component caused by the bound $\sigma$-electrons and the $eQq_{0\textnormal{|el}}$ term taking into account then the unpaired valence electrons. 
The dipole magnetic hyperfine value $c$ from $^{47}$TiO (see Tab.~\ref{tab:measurment_hyper}) can be used to determine the contribution of the valence electrons on the $eQq_0$ \cite{Ryzlewicz1982},
\begin{equation}
eQq_{0\textnormal{|el}}= -2 \cdot \frac{2}{3}\cdot\frac{e^2Q}{\epsilon_0 \mu_B \mu_N\mu_0}\cdot c.
\end{equation} 
For $^{47}$TiO the electric quadrupole parameter $eQq_0$ is -54.612(10)\,MHz.  
The valence electrons contributes with a value of $89.2(7)$\,MHz to the electric quadrupole parameter. 
Thus, the core polarization has to be negative  with $eQq_{0\textnormal{|pol}}=-143.8(7)$\,MHz. 
This large polarization value, roughly three times larger than $eQq_0$, indicates that the closed-shell $\sigma$-electrons are strongly effected by the unbound $\sigma$ electron. The unbound $\delta$ electron seems to compensate the electrostatic polarization effect of the particular values.
By taking a look at the ScO and VO molecules, see Table \ref{tab:dipole_re}, one notices that with increasing atomic number the absolute value of the electric quadrupole coupling decreases. 
Evaluation of the effect of the valence electrons  $eQq_{0\textnormal{|el}}$ for ScO ($19.755(2)$\,MHz), TiO, and VO($-22,6(1)$\,MHz) shows that a compensation of  $eQq_{0\textnormal{|pol}}$  with respect to the $\sigma$-like orbitals takes place. 
The compensation increases with increasing $1\delta$ occupation thereby leading to smaller $eQq_0$ values. 
Comparing the electric quadrupole coupling parameter $eQq_{0}$ of TiO with those of ZrO ($eQq_0=150.5499(46)$\,MHz\added[id=A]{, $Q$=-0.176\,b}) and HfO ($eQq_0=-5952.865(11)$\,MHz\added[id=A]{, $Q$=3.365\,b})\added[id=A]{ after division by the quadrupole moment}, it can be seen that \replaced[id=A]{this quantity directly reflects the electric field gradient at the coupling nucleus\cite{Gordy1984}. The increasing field gradients from TiO over ZrO to HfO is in qualitative agreement with the steeper potential of the corresponding atoms caused by the larger proton number	}{the latter is by an order of magnitude larger, which is due to the increased charge density caused by the increased number of electrons}.  
%\subparagraph{Hyperfine Parameter/Molecular Bond}
\begin{table*}
	\caption{Comparison of hyperfine parameter for $^{47}$TiO and $^{49}$TiO (all values in MHz).\label{tab:measurment_hyper}}	
	\begin{adjustbox}{totalheight=\textheight-2\baselineskip, max width=0.99\textwidth} 
	%\begin{center}
	\begin{threeparttable}
		\begin{tabular}{*{1}{l}*{7}{D{.}{.}{7.7}}}
			\toprule
			&\multicolumn{3}{c}{$^{49}$TiO}&\multicolumn{4}{c}{$^{47}$TiO}\\%\cmidrule(lr){2-4}\cmidrule(lr){5-8}
			&\multicolumn{1}{c}{This work}&\multicolumn{2}{c}{Prev. work}&\multicolumn{1}{c}{This work}&\multicolumn{3}{c}{Prev. work}\\\cmidrule(lr){3-4}\cmidrule(lr){6-8}
			&\multicolumn{1}{c}{Fit A}&\multicolumn{1}{c}{Re-fitted\tnote{a}}&\multicolumn{1}{c}{Exp. work\tnote{a}}&\multicolumn{1}{c}{Fit A}&\multicolumn{1}{c}{Re-fitted\tnote{a}}&\multicolumn{1}{c}{Exp. work\tnote{a}}&\multicolumn{1}{c}{Theo work\tnote{b}}\\%\hline
			\midrule 
			%&&&&&&&\\
			$a$&-53.171(43)&-53.184(44)&-53.155(85)&-53.155(43)&-53.175(50)&-53.36(28)&\multicolumn{1}{c}{-50}\\
			$\Delta a$&-12.585(84)&-12.584(85)&-19.01(36)&-12.647(84)&-12.645(99)&-18.28(67)&\\
			$b$&-259.91(24)&-259.93(28)&-232.94&-259.95(28)&-259.83(24)&-235.3(2.1)&-343.1\\
			$c$&28.10(24)&28.09(22)&\multicolumn{1}{c}{-}&28.09(19)&27.75(34)&\multicolumn{1}{c}{-}&21.3\tnote{d}\\
			$b+c$&-231.813(83)\tnote{c}&-231.840(81)&-231.94(17)&-231.744(83)\tnote{c}&-231.790(90)&-232.19(46)&\\
			%$b_D$&-0.00042(64)&-0.00043(64)&\multicolumn{1}{c}{-}&\\
			$c_D$&-0.0986(29)&-0.0987(31)&\multicolumn{1}{c}{-}&-0.0996(30)&-0.0983(39)&\multicolumn{1}{c}{-}&\\
			%$(b+c)_D$&-0.1003(35)\tnote{d}&-0.09977(35)&\multicolumn{1}{c}{-}&-0.1013(36)\tnote{d}&-0.0970(45)&-0.0178(15)&\\
			$eQq_0$&-44.666(8)&-44.734(12)&-44.72(10)&-54.612(10)&-54.591(7)&-54.587(86)&\\
			\bottomrule
		\end{tabular}
		\begin{tablenotes}
			\footnotesize
			\item [a] Values taken from Lincowski\etal\cite{Lincowski2016}, also $(b+c)_D=-0.0178(15)\,$MHz\\
			\item [b] Values taken from Fletcher\etal\cite{Fletcher1993}\\
			\item [c] Correlation coefficient: $-0.953074$
			%\item [d] Correlation coefficient: $0.414600$\\
			\item [d] Assuming pure $3d_{\delta}$ electron from Ti\\
		\end{tablenotes}
	\end{threeparttable}
	%\end{center}
\end{adjustbox}
\end{table*}
\begin{table*}
	\caption{Atomic, ionic, and mixed parameters$^\text{d}$ used to characterize TiO within the FAC method.\label{tab:facm}}	 
	\begin{center}
	\begin{threeparttable}
		\begin{tabular}{*{1}{l}*{3}{D{.}{.}{4.5}}}
			\toprule
			&\multicolumn{1}{c}{Ti \tnote{a}}&\multicolumn{1}{c}{Ti$^+$\tnote{b}}&\multicolumn{1}{c}{TiO\tnote{c}}\\%\hline
			\midrule 
			%&&&&&&&\\
			$\zeta_{\textrm{3d}}$     [cm$^{-1}$] &82.4(50)&\multicolumn{1}{c}{111(3)}&101.3(37)\\
			$a_{\textrm{3d}}^{01}$ [MHz]  &-43.7(18)&-58.5(55)&-53.5(39)\\
			$a_{\textrm{3d}}^{12}$ [MHz]  &-39.9(37)&-47.1(48)&-44.7(35)\\
			$a_{\textrm{3d}}^{10}$ [MHz]  &52.3(77)&33.6(28)&39.5(37)\\
			$a_{\textrm{4s}}^{10}$ [MHz] &\multicolumn{1}{c}{$-$487(9)}&\multicolumn{1}{c}{$-$836(10)}&\multicolumn{1}{c}{$-$718(33)}\\
			\bottomrule
		\end{tabular}
		\begin{tablenotes}
		% \singlespacing
			\footnotesize
			\item [a] Value taken from Aydin \textit{et al.}\cite{Aydin1989} with configuration $(3d^34s)$\\
			\item [b] Value taken from Bouazza \cite{Bouazza2013}\\
			\item [c] Scaled by the derived ionic character $c_\textnormal{ion} = 0.813(56)$
			\item [d] Spin-orbit coupling ($\vec{L}\vec{S}$): $\zeta_{\textrm{xy}}$; nuclear spin-orbit coupling ($\vec{I}\vec{L}$): $a_{\textrm{xy}}^{01}$; spin-dipole coupling ($I_zS_z$): $a_{\textrm{xy}}^{12}$; Fermi-contact term ($\vec{I}\vec{S}$): $a_{\textrm{xy}}^{10}$
		\end{tablenotes}
	\end{threeparttable}
	\end{center}
\end{table*}
\\

For the first time, a complete mass independent Dunham-type analysis of titanium monoxide for all stable titanium isotopologues has been performed. 
The accuracy of the measurements on six isotopologues allows for a detailed analysis of the influence of the atomic masses on the bond lengths. 
From the molecular hyperfine parameters of the odd-numbered titanium isotopes ($^{47}$Ti ,$^{49}$Ti) a detailed description of the electronic structure of TiO can be given based on measured spectroscopic parameters. 
The mass-invariant parameter set of TiO has been used to calculate molecular parameters for the ground state $X\,^3\Delta$ of $^{44}$TiO. 
As a result from this Dunham-like analysis the line prediction accuracy of $^{44}$TiO is better than 0.1 MHz in the frequency range below 400\,GHz, see Tab.~\ref{tab:measurment_44}.  
\added[id=A]{Models of a core collapse event of a massive star (25 \(\textup{M}_\odot\)) predict isotopic mass ratio of $\nicefrac{^{48}Ti}{^{44}Ti}\approx 3$ \cite{Meyer1995}. If rotational transitions of the most abundant TiO isotopologue are observable in the environment of a SNR, then $^{44}$TiO will also be provided that the age of the object is in the range of the half life of $^{44}$Ti which is 60 years \cite{Ahmad2006}. For objects like SN1987A, a supernova remnant observed in 1987, these conditions may be fulfilled, whereas for the SNR Cas A, which occurred 335 years ago\cite{Fesen2006}, the depletion of $^{44}$TiO is significant and the isotopic ratio $\nicefrac{^{48}Ti}{^{44}Ti}$ may be increased by factor of 50, excluding a strong observation of $^{44}$TiO in this stellar object.} \replaced[id=A]{Nevertheless, accurate rotational transition predications of rare isotopologues, like $^{44}$TiO and $^{26}$AlO \cite{Breier2018}, foster }{However, with this kind of accuracy} a dedicated astronomical search for these rare molecules\added[id=A]{, which may serve as a clock to determine the age of SNRs} \deleted[id=A]{is feasible}.

\begin{table*}
	\caption{Measured rotational transition frequencies (in MHz) of the $^{3}\Delta$ ground state of $^{46,48,50}$TiO. Experimental uncertainties are given in parentheses with $1\sigma$ deviation\cite{ERROR}. The weighted rms value is unitless.\label{tab:measurment_even}}	 
	\resizebox{\linewidth}{!}{%
		{\footnotesize
			\begin{threeparttable}
				\begin{tabular}{*{2}{l}*{1}{D{.}{.}{7.9}}*{1}{D{.}{.}{2.3}}*{1}{D{.}{.}{7.9}}*{1}{D{.}{.}{2.7}}*{1}{D{.}{.}{7.9}}*{1}{D{.}{.}{2.3}}}
					\toprule
					\textit{J$^{\,\prime\prime}$}&$\Omega^{\,\prime\prime}$&\multicolumn{1}{c}{$^{46}$TiO}& \multicolumn{1}{c}{o.-c.}&\multicolumn{1}{c}{$^{48}$TiO}&\multicolumn{1}{c}{o.-c.}&\multicolumn{1}{c}{$^{50}$TiO}&\multicolumn{1}{c}{o.-c.}\\%\hline
					\midrule 
					%&&&&&&&\\
					$\nu=0$&&&&&&&\\
					7&1&255943.678(50)&0.061&253229.064(50)&0.015&250728.856(50)&-0.028\\
					&2&258927.037(100)&0.023&256149.497(50)&-0.024&253592.150(150)&0.157\\
					&3&261511.461(100)&0.080&258677.634(50)&0.050&256068.870(50)&0.085\\
					8&1&287929.144(50)&-0.001&284875.296(50)&-0.043&282062.428(150)&-0.300\\
					&2&291280.974(100)&0.027&288156.474(50)&-0.058&285279.626(150)&0.072\\
					&3&294185.667(100)&-0.003&290998.150(50)&0.123&288063.627(100)&0.156\\
					9&1&319912.095(100)&0.061&316519.049(50)&0.028&313393.923(50)&-0.077\\
					&2&323630.596(100)&-0.073&320159.464(50)&0.037&316963.264(150)&0.179\\
					&3&326854.831(150)&0.025&323313.533(50)&0.082&320053.306(150)&0.049\\
					10&1&351891.923(100)&-0.061&348159.793(50)&-0.005&344722.258(50)&-0.138\\
					&2&355975.594(100)&-0.123&352157.711(50)&-0.038&348642.189(150)&0.053\\
					&3&359518.522(150)&0.299&355623.365(50)&-0.063&352037.535(50)&-0.070\\
					11&1&383868.657(50)&-0.034&379797.325(50)&-0.045&376047.589(100)&-0.033\\
					&2&$\nodata$&&384151.007(50)&-0.038&$\nodata$&\\
					weighted rms&&&0.878&&1.119&&1.318\\
					\midrule
					
					$\nu=1$&&&&&&&\\
					8&2&$\nodata$&&286512.767(100)&-0.008&$\nodata$&\\
					9&2&$\nodata$&&318332.981(100)&0.014&$\nodata$&\\
					weighted rms&&&&&0.353&&\\
					\bottomrule
				\end{tabular}
				\begin{tablenotes}
					\footnotesize
				\end{tablenotes}
			\end{threeparttable}
		}}
	\end{table*}
	\begin{table*}
		\caption{Measured rotational transition frequencies (in MHz) of the $^{3}\Delta_1$ ground state of $^{48}$Ti$^{18}$O. Experimental uncertainties are given in parentheses with $1\sigma$ deviation\cite{ERROR}. The weighted rms value is unitless.\label{tab:measurment_18even}}	 
%		\resizebox{\linewidth}{!}{%
			{\footnotesize
				\begin{threeparttable}
					\begin{tabular}{*{2}{l}*{1}{D{.}{.}{7.9}}*{1}{D{.}{.}{2.3}}}
						\toprule
						\textit{J$^{\,\prime\prime}$}&$\Omega^{\,\prime\prime}$&\multicolumn{1}{c}{$^{48}$Ti$^{18}$O}&\multicolumn{1}{c}{o.-c.}\\%\hline
						\midrule 
						%&&&&&&&\\
						8&1&261372.541(100)&-0.030\\
						9&1&290405.742(100)&0.030\\
						10&1&319436.208(100)&-0.026\\
						11&1&348463.857(100)&-0.012\\
						12&1&377488.377(100)&0.031\\
						weighted rms&&&0.268\\
						\bottomrule
					\end{tabular}
					\begin{tablenotes}
						\footnotesize
					\end{tablenotes}
				\end{threeparttable}
			}
%			}
		\end{table*}
\begin{table}
	\caption{Measured rotational transition frequencies (in MHz) of the $^{3}\Delta$ ground state of $^{47}$TiO and $^{49}$TiO. Experimental uncertainties are given in parentheses with $1\sigma$ deviation\cite{ERROR}. The weighted rms value is unitless.\label{tab:measurment_odd}}	 
%	\resizebox{0.75\linewidth}{!}{%
\begin{adjustbox}{totalheight=\textheight-2\baselineskip, max width=0.5\textwidth}
		%{\tiny
			\begin{threeparttable}
				\begin{tabular}{*{2}{l}*{1}{D{.}{.}{2.1}}*{1}{D{.}{.}{6.7}}*{1}{D{.}{.}{2.3}}*{1}{D{.}{.}{6.7}}*{1}{D{.}{.}{2.3}}}
					\toprule
					\textit{J$^{\,\prime\prime}$}&$\Omega^{\,\prime\prime}$&\multicolumn{1}{c}{\textit{F}$^{\,\prime\prime}$}&\multicolumn{1}{c}{$^{47}$TiO}& \multicolumn{1}{c}{o.-c.}&\multicolumn{1}{c}{$^{49}$TiO}&\multicolumn{1}{c}{o.-c.}\\%\hline
					\midrule 
					%&&&&&&&\\
					7&1&10.5&&&251935.903(100)&0.055\\
					&&9.5&254544.359(100)&0.072&251942.371(150)&0.120\\
					&&8.5&254550.242(100)&0.088&251948.011(150)&0.032\\
					&&7.5&254555.313(150)&0.023&251953.253(150)&0.121\\
					&&6.5&254559.961(150)&0.048&251957.979(100)&0.186\\
					&&5.5&254564.283(150)&0.088&251962.224(150)&0.198\\
					&&4.5&254568.345(150)&0.082&251965.898(150)&0.017\\
					&&3.5&&&251969.486(150)&0.099\\
					&2&3.5&&&$\nodata$&\\
					&&4.5&257495.492(150)&0.090&$\nodata$&\\
					&&5.5&257497.142(150)&0.098&$\nodata$&\\
					&&6.5&257500.228(150)&0.217&$\nodata$&\\
					&&7.5&257504.545(150)&0.300&$\nodata$&\\
					&&8.5&257509.880(150)&0.219&$\nodata$&\\
					&&9.5&257516.287(150)&0.146&$\nodata$&\\
					&&10.5&&&$\nodata$&\\
					8&1&11.5&&&283423.732(100)&-0.040\\
					&&10.5&286357.289(100)&0.047&283429.209(100)&-0.052\\
					&&9.5&286362.415(100)&0.052&283434.410(100)&0.149\\
					&&8.5&286367.102(100)&0.156&283438.860(150)&0.012\\
					&&7.5&286371.218(100)&0.057&283443.147(150)&0.059\\
					&&6.5&286375.267(150)&0.118&283447.094(150)&0.059\\
					&&5.5&286379.134(150)&0.117&283450.880(100)&0.150\\
					&&4.5&&&283454.325(100)&0.125\\
					&2&4.5&&&$\nodata$&\\
					&&5.5&$\nodata$&&$\nodata$&\\
					&&6.5&289675.950(150)&0.086&$\nodata$&\\
					&&7.5&289678.518(150)&0.176&$\nodata$&\\
					&&8.5&289681.959(150)&0.237&$\nodata$&\\
					&&9.5&289686.450(500)&0.531&$\nodata$&\\
					&&10.5&289691.006(150)&0.184&$\nodata$&\\
					&&11.5&&&$\nodata$&\\
					9&1&12.5&&&314908.843(100)&0.072\\
					&&11.5&318167.302(100)&-0.040&314913.733(100)&0.094\\
					&&10.5&318171.990(100)&0.042&314918.150(100)&0.013\\
					&&9.5&318176.329(150)&0.185&314922.381(100)&0.051\\
					&&8.5&318180.056(150)&-0.012&314926.379(100)&0.109\\
					&&7.5&318183.917(150)&0.082&314930.155(100)&0.155\\
					&&6.5&318187.561(150)&0.026&314933.611(100)&0.055\\
					&&5.5&&&314937.085(150)&0.122\\
					&2&5.5&&&$\nodata$&\\
					&&6.5&321848.196(150)&-0.230&$\nodata$&\\
					&&7.5&321849.698(150)&-0.098&$\nodata$&\\
					&&8.5&321851.898(150)&0.003&$\nodata$&\\
					&&9.5&321854.838(150)&0.179&$\nodata$&\\
					&&10.5&321858.242(150)&0.233&$\nodata$&\\
					&&11.5&321862.056(150)&0.210&$\nodata$&\\
					&&12.5&&&$\nodata$&\\
					10&1&13.5&&&346390.745(100)&0.085\\
					&&12.5&349974.347(100)&-0.018&346395.133(50)&0.048\\
					&&11.5&349978.609(150)&0.009&346399.297(50)&0.073\\
					&&10.5&349982.608(150)&0.094&346403.186(100)&0.057\\
					&&9.5&349986.306(150)&0.086&346406.944(100)&0.100\\
					&&8.5&349989.799(150)&-0.016&346410.456(100)&0.050\\
					&&7.5&349993.386(150)&0.010&346414.006(100)&0.162\\
					&&6.5&&&346417.076(150)&-0.105\\
					11&1&14.5&&&377869.209(50)&-0.001\\
					&&13.5&381778.101(100)&0.044&377873.306(100)&0.000\\
					&&12.5&381782.089(100)&0.074&377877.209(100)&0.032\\
					&&11.5&381785.732(100)&0.018&377880.886(100)&0.021\\
					&&10.5&381789.331(100)&0.079&377884.429(100)&0.022\\
					&&9.5&381792.822(100)&0.112&377887.907(100)&0.073\\
					&&8.5&381796.210(100)&0.056&377891.287(100)&0.116\\
					&&7.5&&&377894.463(100)&0.024\\
					\multicolumn{3} {l} {weighted rms}&&0.999&&0.920\\
	
					\bottomrule
				\end{tabular}
				\begin{tablenotes}
					\footnotesize
				\end{tablenotes}
			\end{threeparttable}
		%}
		%}
	\end{adjustbox}
	\end{table}

\begin{table}
	
	\caption{Mass-invariant molecular parameters based on the analysis of six stable titanium isotopolgues,$^{46-50}$Ti$^{16}$O and $^{48}$Ti$^{18}$O.\label{tab:parameter}}	
	\begin{adjustbox}{totalheight=\textheight-2\baselineskip, max width=0.5\textwidth}
		%\footnotesize
		\begin{threeparttable}
			\begin{tabular}{*{1}{l}*{4}{D{.}{.}{3.9}}*{1}{l}}
				\toprule
				%&\multicolumn{3}{c}{This work}&&\\\cmidrule{2-4}
				%Parameter&\multicolumn{1}{c}{This work}&\multicolumn{1}{c}{This work}&\multicolumn{1}{c}{Exp work\tnote{a}}&\multicolumn{1}{c}{Exp work\tnote{a}}&Units\\
				&\multicolumn{2}{c}{This work}&\multicolumn{2}{c}{Prev. work}&\\\cmidrule(lr){2-3}\cmidrule(lr){4-5}
				Parameter\tnote{a}&\multicolumn{1}{c}{Fit A\tnote{b}}&\multicolumn{1}{c}{Fit B\tnote{c}}&\multicolumn{1}{c}{Re-Fitted}&\multicolumn{1}{c}{published\tnote{d}}&Units\\%\hline
				\midrule 
				%&&&&&&&\\
				&&&&&\\
				X\,$^{3}\Delta$ state&&&&&\\
				$U_{00}$&0.0&0.0&0.0&0.0&cm$^{-1}$\\
				$U_{10}\cdot 10^{-3}$&3.4949966(26)&3.4949959(26)&3.4949984(29)& 3.4949989(23)& cm$^{-1}$\,u$^{1/2}$\\
				$U_{20}\cdot 10^{-1}$&-5.47093(45)&-5.47063(46)&-5.47110(46)& -5.47129(36)& cm$^{-1}$\,u\\
				$U_{30}\cdot 10^1$&-1.587(22)&-1.596(22)&-1.582(22)&-1.571(18)& cm$^{-1}$\,u$^{3/2}$\\
				$U_{01}$&6.4227037(25)&6.4234987(74)&6.4207505(16)& 6.4207506(20)& cm$^{-1}$\,u\\
				$\Delta_{U01}^{Ti}$&-8.253(24)&-8.282(25)&-&-&\\
				$\Delta_{U01}^{O}$O&-6.112(8)&-9.722(29)&-&-&\\
				$U_{11}\cdot 10^1$&-1.255825(44)&-1.255631(69)&-1.25598(11)& -1.25599(15)& cm$^{-1}$\,u$^{3/2}$\\
				$U_{21}\cdot 10^3$&-1.3467(57)&-1.3542(61)&-1.3399(71)& -1.325(13)& cm$^{-1}$\,u$^{2}$\\
				$U_{02}\cdot 10^5$&8.67195(38)&8.67186(38)&8.6636(18)& 8.6647(13)& cm$^{-1}$\,u$^{2}$\\
				$U_{12}\cdot 10^6$&1.7341(67)&1.7469(81)&1.7198(120)& 1.701(32)& cm$^{-1}$\,u$^{5/2}$\\
				$U_{03}\cdot 10^{10}$&1.95(17)&2.05(17)&1.17(24)& 1.19(28)& cm$^{-1}$\,u$^{3}$\\
				$A_{00}\cdot 10^{-1}$&5.065030(11)&5.064254(14)&5.065029(11)& 5.065041(10)& cm$^{-1}$\\
				$A_{10}\cdot 10^{3}$&6.77(73)&4.93(84)&6.74(73)& 6.70(58)& cm$^{-1}$\,u$^{1/2}$\\
				$A_{20}\cdot 10^{3}$&-10.61(51)&-10.88(53)&-10.57(51)& -10.10(42)& cm$^{-1}$\,u\\
				$A_{01}\cdot 10^{4}$&-1.797(10)&1.722(56)&-3.1136(36)& -3.1080(32)& cm$^{-1}$\,u\\
				$\Delta_{A01}^{Ti}\cdot 10^{-4}$&6.429(85)&-&-&\\
				$A_{11}\cdot 10^{5}$&-4.25(22)&7.3(28)&-4.28(22)& -4.58(22)& cm$^{-1}$\,u$^{3/2}$\\
				$\gamma_{01}\cdot 10^{2}$&-&9.00(10)&-&-& cm$^{-1}$\,u\\
				$\gamma_{11}\cdot 10^{2}$&-&2.36(52)&-&-& cm$^{-1}$\,u$^{3/2}$\\
				$\lambda_{00}$&1.74974(16)&1.74584(18)&1.74970(16)& 1.74991(14)& cm$^{-1}$\\
				$\lambda_{10}\cdot 10^{2}$&-1.81(11)&-2.00(12)&-1.84(12)& -1.88(9)& cm$^{-1}$\,u$^{1/2}$\\
				$\lambda_{20}\cdot 10^{3}$&3.19(85)&3.13(85)&3.37(86)& 3.60(68)& cm$^{-1}$\,u\\
				$\lambda_{01}\cdot 10^{6}$&6.64(15)&-47.76(44)&7.81(27)& 8.06(19)& cm$^{-1}$\,u$^{3/2}$\\
				$a_{00}\cdot 10^{3}$&5.6220(45)&5.6133(46)&-&-& cm$^{-1}$\,g$_{N}^{-1}$\\
				%$a_{10}\cdot 10^{4}$&7.2(15)&7.2(15)&-&-& cm$^{-1}$\,g$_{N}^{-1}$\,u$^{1/2}$\\
				${\Delta a}_{00}\cdot 10^{3}$&4.620(31)&4.684(31)&-&-& cm$^{-1}$\,g$_{N}^{-1}$\,u$^{1/2}$\\
				$b_{00}\cdot 10^{2}$&2.7481(30)&2.7653(26)&-&-& cm$^{-1}$\,g$_{N}^{-1}$\\
				$c_{00}\cdot 10^{3}$&-2.971(20)&-3.160(20)&-&-& cm$^{-1}$\,g$_{N}^{-1}$\\
				%$b_{01}\cdot 10^{7}$&4.7(81)&-2.2(71)&-&-& cm$^{-1}$\,g$_{N}^{-1}$\,u\\
				$c_{01}\cdot 10^{4}$&1.257(37)&1.205(38)&-&-& cm$^{-1}$\,g$_{N}^{-1}$\,u\\
				${eQq_0}_{00}\cdot 10^{3}$&-6.0320(11)&-6.0322(11)&-&-& cm$^{-1}$\,b$^{-1}$\\
				\\
				weighted rms&1.00&1.01&1.00&1.10&\\
				unweighted rms&0.00547&0.00550&0.00577&0.0054&cm$^{-1}$\\
				\bottomrule
			\end{tabular}
			\begin{tablenotes}
				\footnotesize
				\item [a] \added[id=A]{Mass-invariant molecular parameters of the electronic states A$^3\Phi$ and B$^3\Pi$ are listed in the supplementary material.}\\
				\item [b] \added[id=A]{Fit A  uses the same parameter set as Ram\etal\cite{Ram1999}.}\\
				\item [c] \added[id=A]{Fit B includes the spin-rotational parameters: $\gamma = \gamma_{01}\mu^{-1}+0.5\gamma_{11}\mu^{-1.5}$}\\
				\item [d] Values taken from Ram \textit{et al.}~\cite{Ram1999} and scaled mass-independent values using $\mu_{^{48}\textnormal{Ti}^{16}\textnormal{O}}=11.99388479(6) \textnormal{u}$. Also reported parameter $U_{22}=5.5(31)\cdot10^{-8}$cm$^{-1}$\,u$^{6}$.\\
			\end{tablenotes}
		\end{threeparttable}
		
	\end{adjustbox}
\end{table}

\begin{table}
	\caption{Predicted rotational transition frequencies (in MHz) of the $^{3}\Delta_1$ ground state of $^{44}$TiO and even titanium isotopologues\tnote{a}.\label{tab:prediction_44}}	 
			\begin{threeparttable}
				\begin{tabular}{*{1}{l}*{1}{D{.}{.}{6.8}}}
					\toprule
					\textit{J$^{\,\prime\prime}$}&\multicolumn{1}{c}{$^{44}$TiO}\\%\hline
					\midrule 
					%&&&&&&&\\
					1&64730.891(3)\\
					2&97095.606(4)\\
					3&129459.442(6)\\
					4&161822.105(7)\\
					5&194183.299(9)\\
					6&226542.728(10)\\
					7&258900.092(12)\\
					8&291255.091(13)\\
					9&323607.420(15)\\
					10&355956.774(16)\\
					11&388302.843(18)\\
					12&420645.312(19)\\
					\bottomrule
				\end{tabular}
				\begin{tablenotes}
					\item [a] Frequency uncertainties are obtained by empirical variation of molecular parameter uncertainties.
				\end{tablenotes}
			\end{threeparttable}
	\end{table}

	\begin{table*}
	\caption{Measured rotational transition frequencies (in MHz) of the ground state of $^{48}$TiO$_2$. Experimental uncertainties are given in parentheses with $1\sigma$ deviation \cite{ERROR}. \added[id=A]{For comparison, the calculated values and uncertainties given in the CDMS are also listed\cite{Endres2016}.}\label{tab:measurment_tio2}}	 
	%		\resizebox{\linewidth}{!}{%
	{\footnotesize
		\begin{threeparttable}
			\begin{tabular}{*{1}{l}*{2}{D{.}{.}{6.7}}}
				\toprule
				\textit{$^{\Delta K}\Delta J_{K_a^{\prime\prime},K_c^{\prime\prime}}(J^{\prime\prime}) $}&\multicolumn{1}{c}{This work}&\multicolumn{1}{c}{CDMS\cite{Endres2016}}\\%\hline
				\midrule 
				%&&&&&&&\\
				$^{r}Q_{5,3}(7)$&251977.114(50)&251977.193(11)\\
				$^{r}R_{2,18}(19)$&283567.854(50)&283567.716(22)\\
				%$^{r}R_{3,17}(19)$&284952.166(42)&284951.310(47)\\
				$^{r}Q_{6,4}(10)$&297418.481(50)&297418.556(10)\\
				$^{r}Q_{6,4}(9)$&297539.720(50)&297539.750(11)\\
				$^{p}R_{1,21}(21)$&297553.693(100)&297553.951(50)\\
				$^{r}Q_{6,2}(8)$&297623.750(50)&297623.702(13)\\
				$^{r}Q_{6,2}(7)$&297678.435(50)&297678.543(16)\\
				\bottomrule
			\end{tabular}
			\begin{tablenotes}
				\footnotesize
			\end{tablenotes}
		\end{threeparttable}
	}
	%			}
\end{table*}

\section{Acknowledgment}
\added[id=A]{The authors thank C. Amiot for providing us with his experimental data.} \added[id=A]{Also the authors thank E. Stachowska for discussing the former reported atomic/ionic titanium hyperfine parameters.} The authors acknowledge funding through the DFG priority program 1573 (\textit{Physics of the Interstellar Medium}) under grants GI 319/3-1, GI 319/3-2, GA 370/6-1, GA 370/6-2, and the University of Kassel through P/1052 Programmlinie \textit{Zukunft}.
\newpage
\section{References}

\bibliography{tio}

\begin{thebibliography}{105}
\bibitem[{ERR(????)}]{ERROR}
???? \bibinfo{note}{To take systematic error into account we used the following
  procedure by linking the experimental frequency uncertainty to the
  signal-to-noise ratios (SNR): SNR$>10$: 50\,kHz, SNR$>5$: 100\,kHz, and
  SNR$<5$: 150\,kHz}.
\end{thebibliography}

%\appendix
\section*{Appendix}
\label{sec:app}
\renewcommand{\thesubsection}{\Alph{subsection}}

\subsection{FACM on ($X^3\Delta$)TiO}
\renewcommand{\theequation}{\thesubsection.\arabic{equation}}
\setcounter{equation}{0}
The free atomic comparison method (FACM) is used to evaluate the electronic ground state of TiO, based on the electronic molecular configuration:
\begin{eqnarray}
	X^{3}\Delta:~(core)(9\sigma)^1(1\delta^+)^1 \qquad\text{with}\qquad (core) = (1\sigma)^2-(8\sigma)^2(1\pi)^4-(3\pi)^4.
	\label{app:1}
\end{eqnarray}
Theoretical predictions from Bauschlicher\etal\cite{Bauschlicher1983} reveal a polarization of the titanium atom. The $9\sigma$ molecular orbital is electrostatically pushed away from the oxygen and mixes with $4p$ and $3d$ atomic orbitals of titanium. Hence, the unpaired molecular orbitals (see Eq.\,(\ref{app:1})) are in good approximation linear combinations of $4s$, $4p$, and $3d$ orbitals of titanium.  The small contribution of the oxygen $2p$ orbitals can be neglected and the $\sigma$ orbital can be expressed as, 
\begin{eqnarray}
	\ket{9\sigma}= c_{4s}^{9\sigma}\ket{4s}-c_{3d}^{9\sigma}\ket{3d_\sigma}-c_{4p}^{9\sigma}\ket{4p_\sigma},\\
	\end{eqnarray}
with the normalization $({c_{4s}^{9\sigma}})^2+({c_{3d_\sigma}^{9\sigma}})^2+({c_{4p_\sigma}^{9\sigma}})^2=1$. The $1\delta$ molecular orbital is well represented by the $3d$ Ti orbital,
\begin{eqnarray}
	\ket{1\delta}=\ket{3d_\delta}.
	\label{eq:delta}	
\end{eqnarray} 
The TiO bond is thought of being partly covalent and ionic character. This assumption is based on the electron donation in TiO \cite{Bauschlicher1983}. The atomic orbital can be described as
\begin{eqnarray}
	\ket{\chi}=c_{atom}^2 \ket{\chi(Ti)}+c_{ion}^2 \ket{\chi(Ti^+)}
	\label{eq:atio}
\end{eqnarray}
utilizing the normalization relation $c_{atom}^2+c_{ion}^2 =1 $.
The FAC method makes use of the atomic fine- and hyperfine-structure parameters. Therefore, this method balances the molecular-orbital properties by the atomic and ionic structure parameters. In the further description the parameters derived from Eq.(\ref{eq:atio}) are called `mixed parameter values'. In this paper, we use the model-space parameter description  $(3d+4s)^{N+2}$ of Ti $(N=2)$ \cite{Aydin1989}, or Ti$^+$ $(N=1)$ \cite{Bouazza2013}. Where the one-body-parameter $a^{\kappa k}_{nl}(l^{N+M}s^{2-M})$ are related to the model-space parameters $a^{\kappa k}_{nl}$,
\begin{eqnarray}
	a^{\kappa k}_{nl}(l^{N+M}s^{2-M})=a^{\kappa k}_{nl}+\frac{2}{2l+1}(1-N-M)a_1-\sqrt{\frac{2}{2l+1}}(2-M)a_4+\frac{2}{2l+1}a_5\delta(M,0),
	\label{eq:at1}
	\end{eqnarray}
for $\kappa k =	01,12$. In the case of $\kappa k =	10$ it follows:
\begin{eqnarray}
a^{10}_{nl}(l^{N+1}s^{1})&=&a^{10}_{nl}-\frac{2}{2l+1}a_9,\nonumber\\
a^{10}_{nl}(l^{N+2})&=&a^{10}_{nl}-\frac{2}{2l+1}(a_{10}-a_9).
\label{eq:at2}
\end{eqnarray}
In the case of atomic Ti and ionic Ti$^+$ we use the electronic configuration $(3d^3s^1)$ and $(3d^2s^1)$, respectively, within the model space. 
It represents the donation of unbound TiO electrons originating from the Ti atom. 
Therefore in case of titanium, the Eq.\,(\ref{eq:at1}) and (\ref{eq:at2}) reduce to the following expressions for the atomic/ionic parameters
\begin{eqnarray}
a^{\kappa k}_{3d}(3d^{3}4s^{1})&=&a^{\kappa k}_{3d}-\frac{4}{5}a_1,\\
a^{10}_{3d}(3d^{3}4s^{1})&=&a^{10}_{3d}-\frac{2}{5}a_9,
\end{eqnarray}
and in case of Ti$^+$ we write 
\begin{eqnarray}
a^{\kappa k}_{3d}(3d^{2}s^{1})&=&a^{\kappa k}_{3d}-\frac{2}{5}a_1,\\
a^{10}_{3d}(3d^{3}s^{1})&=&a^{10}_{3d}-\frac{2}{5}a_9.
\end{eqnarray}
First, we want to determine the ionic character for the ground state $X^3\Delta$ of TiO.
A convenient way to calculate the ionic character is by using the molecular hyperfine nuclear spin-orbit coupling parameter $a$, as defined by
\begin{equation}
a= 2 \mu_Bg_N\mu_N\frac{\mu_0}{4\pi}\frac{1}{\Lambda}\bra{\Lambda\Sigma}\sum_i\frac{\hat{l}_{zi}}{r_i^3}\ket{\Lambda\Sigma},
\end{equation} 
with the electron Bohr magneton $\mu_B$, the nuclear Bohr magneton $\mu_N$, the nuclear \textit{g} factor $g_N$, the magnetic constant $\mu_0$ and the z component of the one-electron orbital angular operator $\hat{l}_{zi}$. The expectation value from a $\sigma$-type orbital is zero.  Therefore only the $1\delta$ electron of TiO in the electronic ground state will contribute to the nuclear spin-orbit interaction: $a = a(1\delta)$. This leads to
\begin{eqnarray}
	a=a(1\delta)&=&c_{atom}^2 \overbrace{a_{3d_\delta}(Ti)}^{a_{3d}^{01}(Ti)}+c_{ion}^2 \overbrace{a_{3d_\delta}(Ti^+)}^{a_{3d}^{01}(Ti^+)}\label{eq:a_delta}\\
	&=&(1-c_{ion}^2) {a_{3d}^{01}(Ti)}+c_{ion}^2 {a_{3d}^{01}(Ti^+)}\\
	\rightarrow c_{ion}^2 &=&\frac{a -  {a_{3d}^{01}(Ti)}}{ {a_{3d}^{01}(Ti^+)}- {a_{3d}^{01}(Ti)}}\\
	&=&0.638(242)
\end{eqnarray}
Another way to calculate the ionic character is to make use of the molecular fine structure, by using the spin-orbit parameter $A$,
\begin{eqnarray}
	A=\frac{1}{\Lambda\Sigma}\bra{\Lambda\Sigma}\sum_i \hat{a}_i\hat{l}_{zi}\ket{\Lambda\Sigma}, \label{app:sin_orbit}
\end{eqnarray}
which for TiO results in,
\begin{eqnarray}
	A&=&\nicefrac{1}{2}\bra{1\delta}\hat{a}_\delta \hat{l}_z\ket{1\delta}\\
	\textnormal{using Eq.\,(\ref{eq:delta})\,\&\,(\ref{eq:a_delta})}\rightarrow 2A&=&c_{atom}^2 \overbrace{A_{3d_\delta}(Ti)}^{\zeta_{3d}(Ti)}+c_{ion}^2 \overbrace{A_{3d_\delta}(Ti^+)}^{\zeta_{3d}(Ti^+)}.
\end{eqnarray}
Furthermore, we use the atomic Ti spin-orbit parameter $\zeta_{3d}(Ti)=\zeta_{3d}(3d^34s^1)=82.4(50)\textnormal{\,cm}^{-1}$ \cite{Stachowska1997} and the ionic Ti with $\zeta_{3d}(Ti^+)=\zeta_{3d}(3d^24s^1)=111(30)\textnormal{\,cm}^{-1}$\cite{Bouazza2013} within our FAC method. Determining the ionic character results in a value of
\begin{eqnarray}
c_{ion}^2&=&\frac{2A-\zeta_{3d}(Ti)}{\zeta_{3d}(Ti^+)-\zeta_{3d}(Ti)}\\
&=&0.661(91).
\end{eqnarray}
The ionic character of TiO derived from the fine-structure parameter $A$ with a value of $0.661(91)$ is in excellent agreement with the value extracted from the hyperfine parameter $a$ with
$c_{ion}^2$ = $0.638(242)$. Comparing this result with the previously attained value using the hyperfine structure parameter $a$ reveals that due to the large error of the ionic spin-orbit coupling  parameter $a_{3d}^{10}(Ti^+)$ a larger uncertainty of $c_{ion}$ follows. 
Our simple FA method is in good agreement with the value derived from Namiki\etal\cite{Namiki2004} with $c_{ion}^2$ = $0.623(6)$. Namiki\etal evaluate the relative vibronic intensities of different optical transitions from their TiO measurements and determine the orbital characters via the electronic transition moment of TiO. They underestimate the uncertainties by taking only the molecular parameters with their uncertainties into account. It seems that the work by Namiki\etal makes no use of the error for their derived atomic or ionic parameters within the FAC method\cite{Namiki2003}. In following part we want to calculate the $\sigma$ character of TiO by making use of our ionic value $c_{ion}^2=0.661(91)$  derived from the fine-structure constant $A$. \\ 
The molecular Fermi-contact parameter $b_F$ is composed of the nuclear spin-electron spin interaction parameter $b$ and the dipolar parameter $c$,
\begin{eqnarray}
	b_F=b+\nicefrac{c}{3}.
\end{eqnarray}
The Fermi-contact interaction is very sensitive to the $s$-character of the unpaired electron. All unpaired electrons have an unquenched orbital angular momentum. Especially, the spin polarization induced by the exchange energy between unpaired and $\sigma$-bonding electrons gives a small and negative $b_F$ contribution. However, the Fermi interaction of non $s$-orbitals is expected to cancel out each other and thus to vanish.
\begin{eqnarray}
	b_F&=& 2 g_S\mu_Bg_N\mu_N\frac{\mu_0}{3}\frac{1}{\Sigma}\bra{\Lambda\Sigma}\sum_i \hat{s}_{zi} \delta(r)\ket{\Lambda\Sigma}\\
	b_F&=&\nicefrac{1}{2}\left(b_F(9\sigma)+b_F(1\delta)\right)\\
	b_F(9\sigma)&=& {c_{4s}^{9\sigma}}^2 \overbrace{\biggl(c_{atom}^2 \underbrace{{b_{F,4s}(Ti)}}_{a_{4s}^{10}(Ti)}+c_{ion}^2 \underbrace{{b_{F,4s}(Ti^+)}}_{a_{4s}^{10}(Ti^+)}\biggr)}^{a_{4s}^{10}(TiO)} + {c_{3d_\sigma}^{9\sigma}}^2 {a_{3d}^{10}(TiO)}+{c_{4p_\sigma}^{9\sigma}}^2{a_{4p}^{10}(TiO)}  \label{app:bF_sigma}\\
	b_F(1\delta)&=&{a_{3d}^{10}(TiO)}\\
	\rightarrow 2b_F&=&{c_{4s}^{9\sigma}}^2 a_{4s}^{10}(TiO)+\left(1+{c_{3d_\sigma}^{9\sigma}}^2\right){a_{3d}^{10}(TiO)}+{c_{4p_\sigma}^{9\sigma}}^2{a_{4p}^{10}(TiO)} 
	\label{app:bF}
\end{eqnarray}
Presuming the $\sigma$ character is purely originated by the $\ket{4s}$ orbital, we simply assume that the atomic $\sigma$ orbitals compensate each other,
\begin{eqnarray}
	2b_F-{a_{3d}^{10}(TiO)}&=&{c_{4s}^{9\sigma}}^2 a_{4s}^{10}(TiO)\\
	\rightarrow{c_{4s}^{9\sigma}}^2&=&0.753(35).
\end{eqnarray}
By comparing our $4s$-$\sigma$ character with the previously derived value of $0.801(2)$ by Namiki\etal\cite{Namiki2004} it can be seen that it is in good agreement. Our purely experimentally derived value can be interpreted as a lower limit, because our compensation assumption of $4p_\sigma$ and $3d_\sigma$ electrons neglects further influence on the $4s$ state. 
By considering this large $\sigma$-character on the $4s$ orbital we assume that the remaining $\sigma$-contribution is distributed equally over the atomic $3d$ and $4p$ orbitals. Considering the large error $err({c_{4s}^{9\sigma}}^2)=0.035$ derived from the experimental measurement, the assumption is reasonable with respect to its absolute value. Applying the normalization rule the $\sigma$ character coefficients are given by 
\begin{eqnarray}
	{c_{3d_\sigma}^{9\sigma}}^2={c_{4p_\sigma}^{9\sigma}}^2&=&0.12(2).
	\end{eqnarray}
These results lead to the same values within error limits as presented by Namiki\etal\cite{Namiki2004} $({c_{3d_\sigma}^{9\sigma}}^2=10\textnormal{,\,}{c_{4p_\sigma}^{9\sigma}}^2=0.11)$.
Now, the mixed Fermi-contact parameter $a_{4p}^{10}(TiO)$ can be determined as 
	\begin{eqnarray}
	{c_{3d_\sigma}^{9\sigma}}^2 {a_{3d}^{10}(TiO)}&=&-{c_{4p_\sigma}^{9\sigma}}^2{a_{4p}^{10}(TiO)},\nonumber\\
	\rightarrow {a_{4p}^{10}(TiO)} &=&-\frac{{c_{3d_\sigma}^{9\sigma}}^2}{{c_{4p_\sigma}^{9\sigma}}^2} {a_{3d}^{10}(TiO)}=-40(9) \textnormal{\,MHz}.
\end{eqnarray}
To summarize,  this value has been determined by the compensation approach, the equally distribution assumption and with the value of atomic $3d$ mixed parameter $a_{3d}^{10}(TiO)$. This gives us a hint how much the atomic $4p$ orbital effects the Fermi-contact parameter $b_\textnormal{F}$ and the $\sigma$ character of the $4s$ orbital. Even in case that  the compensation approach is not fully valid the $4s-\sigma$ character will change only slightly.\\
As a last quantity of the magnetic hyperfine molecular parameter set the dipolar Fermi-contact parameter $c$ is investigated with 
	\begin{eqnarray}
		c&=& \frac{3}{2} g_S\mu_Bg_N\mu_N\frac{\mu_0}{4\pi}\frac{1}{\Sigma}\bra{\Lambda\Sigma}\sum_i \frac{\left(\cos^2\theta_i-1\right)}{r^3}\hat{s}_{z} \ket{\Lambda\Sigma}.\label{app:c} \\
		c&=&\nicefrac{1}{2}\left(c(9\sigma)+c(1\delta)\right)\\
		c(9\sigma)&=&\frac{3}{2} \Biggl({c_{4s}^{9\sigma}}^2 \underbrace{\left<\cos^2\theta-1\right>_{4s}}_{=0}\overbrace{\biggl(c_{atom}^2 \underbrace{{c_{4s}(Ti)}}_{a_{4s}^{12}(Ti)}+c_{ion}^2 \underbrace{{c_{4s}(Ti^+)}}_{a_{4s}^{12}(Ti^+)}\biggr)}^{a_{4s}^{12}(TiO)} \nonumber\\
		&&+ {c_{3d_\sigma}^{9\sigma}}^2  \underbrace{\left<\cos^2\theta-1\right>_{3d_\sigma}}_{=\nicefrac{4}{7}} {a_{3d}^{12}(TiO)}+{c_{4p_\sigma}^{9\sigma}}^2 \underbrace{\left<\cos^2\theta-1\right>_{4p_\sigma}}_{=\nicefrac{4}{5}}{a_{4p}^{12}(TiO)}\Biggr)\\
		c(1\delta)&=&\frac{3}{2}\underbrace{\left<\cos^2\theta-1\right>_{3d_\delta}}_{=-\nicefrac{4}{7}}{a_{3d}^{12}(TiO)}\\
		\rightarrow 2c&=&\underbrace{\frac{6}{7}\left({c_{3d_\sigma}^{9\sigma}}^2-1\right){a_{3d}^{12}(TiO)}}_{2\mathcal{T}(a_{3d}^{12}(\textnormal{TiO}))\approx 33.6(27)\textnormal{\,MHz}}+\underbrace{\frac{6}{5}{c_{4p_\sigma}^{9\sigma}}^2{a_{4p}^{12}(TiO)}}_{2\mathcal{T}(a_{4p}^{12}(\textnormal{TiO}))\approx 23.2(27)\textnormal{\,MHz}}  \label{app:2c}\\
		\rightarrow {a_{4p}^{12}(TiO)}&=&157(29)\textnormal{\,MHz}.
	\end{eqnarray}
 The resulting mixed $4p$ atomic dipolar parameter value $a_{4p}^{12}(TiO)$ is more than three times larger than the $3d$ value $a_{3d}^{12}(TiO)$ and of opposite sign. However, with increasing $\sigma$ character of the $4s$ state and the $\sigma$-compensation approach, the values of $c_{3d_\sigma}^{9\sigma}$ and $c_{4p_\sigma}^{9\sigma}$ decrease, therefore the ${a_{4p}^{12}(TiO)}$ value will also increase. The analysis on the molecular dipolar parameter $c$ reveals that the $\delta$ orbital contribution $\mathcal{T}(a_{3d}^{12}(\textnormal{TiO})$ is roughly half the value derived from the $4\pi_\sigma$ orbital of TiO $\mathcal{T}(a_{4p}^{12}(\textnormal{TiO})$. Therefore, we may conclude that the strong polarization of TiO leads to an increase of molecular dipolar parameter $c$.   
 
% \listofchanges[style=list]
\end{document}